\documentclass[plain]{hvdthesis}
\usepackage{subcaption}

\makeatletter
\providecommand\sf@counterlist{}  
\makeatother

\usepackage[colorlinks=true,
            linkcolor=blue,        
            citecolor=blue,        
            urlcolor=blue,         
           ]{hyperref}

\bibpunct{(}{)}{;}{a}{}{,}

\usepackage{epstopdf}
\usepackage{epsfig}
\usepackage{aasmod}
\usepackage{amsmath}
\usepackage{amssymb}
\usepackage{rotating}
\usepackage{verbatim}
\usepackage{longtable}
\usepackage{lscape}
\usepackage{subfig}
\usepackage{float}
\usepackage{color}
\usepackage{fltpage}
\usepackage{url}
\usepackage{rotating}
\usepackage[innercaption]{sidecap}
\usepackage{pdfpages}
\usepackage{xspace}
\usepackage[letterpaper,margin=0.75in]{geometry}

\usepackage{fancyhdr}
\definecolor{orange}{rgb}{.8,0.4,0}

\author{Aaron Philip }
\title{Extracting Barrier Distributions from Fusion Cross
Sections}
\department{Department of Physics and Astronomy}
\subject{Nuclear Physics Theory}
\month{June}
\year{2025}
\advisor{Dr. Kyle Godbey}

\let\cleardoublepage\clearpage

\begin{document}



\frontmatter

\makecover
\abstract{
Studying fusion cross sections provides insight into the fusion process, details about the internal structure of heavier nuclear systems, and a window into astrophysical processes. It has long been known that a simple model of fusion as the tunneling of a state through a one-dimensional barrier of an effective interaction is not sufficient to capture the full breadth of microscopic dynamics that influence the fusion process in medium and heavy nuclei.

One interpretation of the observed deviations from the 1D tunneling model of fusion is that couplings to collective states and other channels induce a continuous ``distribution of barriers''. Extracting this barrier distribution from the measured cross-sectional data amounts to taking the 2nd derivative of the energy-weighted cross section \citep{Rowley1991}. Barrier distributions are immensely useful for comparing theoretical model predictions and experimental results. In practice, barrier distributions extracted from experimental data are highly sensitive to the quality of collected experimental data and the choice of step size when using standard point difference schemes. Some progress has been made in fitting smooth functions then differentiating or computing quantities less sensitive to noise to characterize the barrier \citep{Scamps2018, Rehm2016}. In a recent work, \cite{Godbey2024} used Gaussian processes to determine a probability distribution of potential continuous barrier distributions that could reasonably describe the data. While this method reduces numerical instability and provides uncertainty estimates, it is not yet clear what the limits of this regression technique are.

In this thesis, we perform an extensive analysis determining how effectively Gaussian processes capture key features of the barrier distribution function as the quality of data varies. We benchmark against realistic simulated data generated from a simple model of fusion excitation functions. We then extend the method to incorporate recently developed Bayesian machine learning inference algorithms \citep{AutoBNN2024} to explore several expressive and interpretable Bayesian Neural Network (BNN) architectures in an automated fashion. We benchmark results on the same toy problem, then use our conclusions to calibrate models to experimental data and extract barrier distributions that can describe the data. We find that Gaussian processes often exhibit aliasing at higher energies of the barrier distribution. We demonstrate that the BNN architectures can more faithfully recover the barrier distribution with quantified uncertainties at all energies, while also identifying key regions of high uncertainty and model discrepancy to determine precisely where additional experiments would be maximally impactful. All methods are robust to data sparsity and irregularity, but we find that the single most important factor dictating the fidelity of all models is the relative size of experimental uncertainties. We release an open-source version of our analysis and a user-friendly implementation of our method to encourage its future usage for experimental analysis. 
}

{\singlespace
\tableofcontents
}
\newpage
\clearpage


\thispagestyle{plain}
\addcontentsline{toc}{chapter}{Acknowledgments}
\chaptermark{Acknowledgments}

\vskip 0.5cm
{\centerline {\Large \bf Acknowledgments}}
\vskip 0.5cm
\normalsize
First, I thank my mentors: Doctors Kyle Godbey, Pablo Giuliani, and Witek Nazarewicz for their mentorship, pedagogy, and patience with me over the last few years. It has been great to work with and learn from you all in many different ways. I hope this thesis reflects the individual impacts you each have had on my scientific outlook. Many thanks to my group mates Pranav Agarwal, Troy Dasher, An Le, Lauren Jin, and Andrew Yeoman-Stephenson for their support and friendship while developing these ideas. Special thanks to Lauren for her excellent feedback on most things I write. And thank you to my family for their support and tolerance and for making me take breaks while writing.



\vspace*{\fill}


\mainmatter
\pagestyle{fancy}


\chapter[Introduction]{Introduction}\label{c:intro}
Atomic nuclei are immensely difficult systems to model. Nuclei are quantum many-body systems held together by the nuclear force which is difficult to probe and non-trivial to treat analytically. Moreover, nuclear systems are multiscale in every sense of the word: collective excitations live in the sub-MeV energy range while binding energies are in the hundreds of MeVs \citep{Sun2025}, radioactive nuclei can have half-lives ranging from $10^{-24}$ s \citep{Kondev2021} to timescales exceeding the age of the universe \citep{Aprile2019}, and the size scale of interest ranges from $10^{-15}$ m when looking at individual nucleons \citep{Kelkar2016} all the way up to kilometers when investigating neutron stars \citep{Huxford2024}. 

In the early days of nuclear science, quantum mechanics did not seem capable of capturing this wide array of systems. To construct the Hamiltonian that solves the time-independent Schroedinger equation: \begin{equation}
    \hat H|\psi\rangle = E |\psi\rangle
\end{equation}
early nuclear physicists had to confront the extremely challenging nuclear force. This force had no clear analytic form and strange properties that had to be inferred from experiment. Among other things, unlike the central Coulomb force, the nuclear force was spin-dependent \citep{Fernbach1955}, finite range \citep{Bethe1949}, and dependent on the angle between two nucleons' spins (the tensor force) \citep{Kellogg1939, Gerjuoy1942}. As such, specifying the system Hamiltonian itself (i.e. stating the problem), was itself nearly impossible.

Even if the Hamiltonian was known exactly, the computer was yet to be invented--quantum mechanics could not feasibly model most systems except for a few limited cases as was the case in electronic theory. Thus, building up a nuclear theory from pure quantum mechanics was not possible nor practical. 

However, a central principle of theoretical physics is that not all degrees of freedom are equally useful. For example, describing the motion of a classical system like a tennis ball does not require a quantum treatment of the dynamics of each constituent atom. Similarly, describing properties of nuclei was still possible, even without starting from a fundamental quantum treatment of all particles composing the nuclei. Historically, nuclear theorists designed phenomenological models and fitted them to experimental data based on the resolution of the system they were studying. Theorists developed new models of the nucleus by examining experimental data, identifying connections across fields, and utilizing theories that could flexibly incorporate new physics as it was discovered. In turn, experimentalists tested these theories, stimulating revisions and additions to existing theoretical frameworks. Our current understanding of nuclear structure has always been driven by the theory-experiment cycle: the process of theorists developing models to explain experimental results, which in turn informs and motivates new experiments.

The following section provides more detail about the historical developments of nuclear science. These developments contextualize and motivate the rest of this work which focuses on contributing to the theory-experiment cycle by improving the extraction of barrier distributions experimental fusion cross section data, utilizing tools from a broad array of disciplines, and proposing a diverse set of models for our data to motivate new experiments. 

\section{Brief Historical Background}
Soon after the neutron's discovery in 1932, Weizsaecker, Bethe, and Bacher proposed the Liquid Drop Model (LDM) in the mid 1930s \citep{Weizacker1935, BetheBacher1936} to explain binding energy data by approximating the nucleus as an incompressible ball of fluid. An improvement to this model of binding energies came in 1949, when Maria Goeppert Mayer drew inspiration from the spin-orbit effect in atomic systems and her shell model of the nucleus was able to capture the experimentally observed ``magic numbers'' of protons and neutrons that result in highly stable isotopes \citep{Mayer1950a, Mayer1950b}. 

In parallel, Yukawa proposed in 1935 that nucleons exchange a massive particle \citep{Yukawa1935} by drawing an analogy to the massless photon exchanged via the electromagnetic interaction in Quantum Field Theory. His proposed particle, the pion, was discovered in the late 1940s \citep{Lattes1947}. 

Motivated by the experimental discoveries of more and more new hadrons beyond just the proton and neutron, the quark model \citep{GellMann1964, Zweig1964} was proposed to explain them, which in turn motivated the search and ultimate discovery of quarks in the 1970s \citep{Friedman1969}. Subsequently, Quantum Chromodynamics (QCD) theory was born. 

While QCD unlocked new fundamental degrees of freedom with which to understand the lightest nuclear systems, heavy nuclei were still out of reach. Density functional theory originated in electronic theory  as a way to compute the many-body electronic wavefunction \citep{HohenbergKohn1964, KohnSham1965}, but was adapted by nuclear theorists in the 1970-80s to construct energy density functionals, enabling the computational treatment of heavy systems \citep{VautherinBrink1972, Negele1972, Gogny1980}. 

In these landmark advancements of nuclear science, there are clear patterns that emerge again and again. Theorists were consistently able to guide experiments forward with phenomenological models even though they did not have a fundamental theory to build their models upon. As such, data-driven scientific modeling has always been a key part of nuclear science. Additionally, advances in nuclear theory were often motivated by drawing connections to other fields like quantum field theory or electronic structure. These are still integral to advancing the field today.

\section{Consequences of Phenomenology}
 However, there are important considerations when we utilize phenomenological models and try to connect different fields: \begin{itemize}
    \item \textbf{Separation of Scale and Expressivity: }
    Even though we may have a more complete understanding of elementary particle physics, this does not mean that choosing QCD degrees of freedom is either necessary or tractable when describing many systems. Instead, we select models by choosing the ``coarsest resolution'' that still allows us to describe the system of interest. However, our models are only as expressive as our assumptions allow them to be. Choosing coordinates for a problem means that we assume finer resolving details do not affect our results much--this is a subjective choice that is challenging to get right. For example, the inclusion of 3N and higher body forces in nuclear \textit{ab initio} theory has proven essential to achieve agreement with experiment, yet a great body of work has gone into working with just two body forces.
    \item \textbf{Calibration and Extrapolation: } 
   Most theoretical models of structure are ultimately calibrated to available experimental data: the LDM, shell models, energy density functionals, optical potentials, Low Energy Constants from $\chi$EFT, etc. No truly \textit{ab initio} theory (in the sense that electronic theory uses the term) exists that can describe most systems. This is especially important when extrapolating to the edges of the nuclear chart with models calibrated to existing experimental data--often, our models have uncontrolled approximations or are missing physics. Fitting available data well does not imply a phenomenological model will perform similarly well when extrapolating \citep{Phillips2021}. 
    \item \textbf{Theoretical Uncertainty Quantification: }
    As a result of the previous point, there is a real need for theoretical uncertainty quantification (UQ) on our model predictions. In nuclear science, theoretical uncertainties tend to dominate experimental ones. In order to have an effective feedback loop between experimentalists and theorists that drives advances forward, theorists must be able to estimate their calculation uncertainties when guiding experimental design and interpreting results.
\end{itemize}

\section{This Work}
This thesis contributes to the theory-experiment cycle by developing techniques to extract information about nuclear structure from experimental fusion cross sections. These contributions live at the intersection of nuclear physics, scientific computing, statistical modeling, and machine learning. This interdisciplinary approach enables us to explore many different possible barrier distributions that describe the same cross sections while also providing interpretable uncertainty estimates. We thoroughly benchmark these methods on realistic toy problems to best understand the contexts in which different models are robust. Moreover, considering an array of diverse and expressive models gives us more confidence in their predictions when models agree and motivates the need for further experiments when we see high discrepancies and uncertainties across models. 
We build on existing work \citep{Godbey2024} that utilized Gaussian processes to extract information about the interactions between nuclei in a principled and statistically sound way. We carry out a comprehensive study of how the method performs in different realistic scenarios by benchmarking it on a toy problem: the Wong formula for fusion cross sections \citep{Wong1973}. We then extend the technique to enable the automated exploration of more sophisticated statistical models by adapting AutoBNN: a new tool released by Google \citep{AutoBNN2024} for training interpretable Bayesian Neural Networks to perform probabilistic time series forecasting. Finally, we use the main conclusions obtained by studying the toy problem to model the same experimental datasets used in the original analysis. 

The rest of the thesis is organized as follows: 
\begin{itemize}
\item In \textbf{Chapter 2}, we introduce the Wong benchmark problem, existing formalism used to interpret fusion cross sections, fundamentals of Gaussian process theory, and the framework for extracting a continuous posterior over potential barrier distributions. We demonstrate how the method performs as the quality of data varies in realistic ranges. We also introduce a method that accounts for scenarios where no quoted experimental uncertainties are available.
\item In \textbf{Chapter 3}, we present a pedagogical introduction to using Bayesian statistics for high-dimensional parameter inference in model regression. This leads into a discussion of Bayesian Neural Networks and the AutoBNN tool. We then present results using AutoBNN, high performance computing, and software development to extract barrier distributions from the benchmark problem introduced in the previous chapter and demonstrate that certain predefined architectures do an exceptional job extracting posterior distributions with well-quantified uncertainties.
\item In \textbf{Chapter 4}, we apply conclusions from the previous chapters to model the experimental fusion data that \cite{Godbey2024} was first applied to. We contribute novel insights into what features the true barrier distribution is likely to have and where more experimental data is needed. We also analyze real experimental data for which no quoted uncertainties are easily available. 
\item In \textbf{Chapter 5}, we report the key takeaways that should inform the use of these tools to analyze other experimental fusion cross sections and discuss outlooks on extending this work. 
\end{itemize}

\clearpage
\chapter{Gaussian Process Regression for Barrier Distributions}\label{c:chap2}
\chaptermark{}

\section*{Abstract}
The Gaussian process Regression (GPR) method in \cite{Godbey2024} for fitting cross sections to extract barrier distributions is advantageous because it is numerically stable and automatically provides estimates of uncertainty. However, it is not yet known how reliable the extracted barriers and the estimated uncertainties are at energies higher than the primary barrier. In this chapter, we seek to fill this gap by benchmarking the GPR method on a toy model: the Wong formula \citep{Wong1973} which describes the collision of spherical (deformed) nuclei using a simple analytic form with 3 (4) parameters. We use the Wong formula to investigate how effectively Gaussian processes can represent the true barrier distribution under realistic measurement errors, data resolution, and data irregularity.

\section{Fusion Cross Sections and Barrier Distributions}
Colliding heavy ions allows us to study the fusion process in detail and extract information about the internal structure of the colliding nuclei. Moreover, we learn about rare and superheavy elements and under what conditions these isotopes can be synthesized \citep{Wilczy2004}--this enables us to form these products on Earth in facilities like FRIB as well as providing information about processes like stellar nucleosynthesis taking place across the universe. 

A typical approach to scattering is to consider a simple central potential that describes the effective interaction of the entire system and try to extract the cross section based on the probability of tunneling through the single barrier arising from the competition of the attractive strong nuclear force and the repulsive Coulomb potential. However, heavy ion systems are far more complex than this and experimental cross section data revealed that one-dimensional single barrier models did not describe the underlying dynamics fully. Couplings to low-lying collective excitations \citep{Hagino2012, Timmers1998}, inelastic or particle transfer channels \citep{Wilczy2004}, and deformations \citep{Dasgupta1998} all play a significant role in the fusion process. 

Capturing these effects was empirically discovered to be possible by considering a continuous distribution of barriers $D(E)$ rather than a single barrier. In 1991, \cite{Rowley1991} drew an important connection by showing that one could extract the barrier distribution function from the data by taking the second derivative of the energy-weighted cross section \begin{equation}
    D(E) = \frac{1}{\pi R^2}\frac{d^2(E\sigma)}{dE^2}
\end{equation} with $\frac{1}{\pi R^2}$ a normalization constant such that the distribution function integrated to unity. Importantly, this meant that one could identify the relevant couplings at play almost directly from the data.

In practice, extracting this distribution function from experimental data is challenging. Standard practice is to use a 3-point difference formula \citep{Dasgupta1998} at energy $E' = (E_1 + 2E_2 +E_3)/4$ given by (dropping the normalization constant)
 \begin{equation}
   D_f(E') = \left. \frac{d^2(E\sigma)}{dE^2} \right |_{E'} \approx 2[\frac{(E\sigma )_3 - (E\sigma)_2}{E_3 - E_2} - \frac{(E\sigma )_2 - (E\sigma)_1}{E_2-E_1}](\frac{1}{E_3 - E_1})
    \end{equation}
with an approximate uncertainty of \begin{equation}
    \delta D_f = (\frac{E}{\Delta E^2})[(\delta \sigma)^2_1+4(\delta \sigma)^2_2+(\delta \sigma)^2_3]^{1/2}
\end{equation}
In principle, the estimation of derivatives will be exact in the limit that the step size $\Delta E \to0$. However, the $\frac{1}{(\Delta E)^2}$ scaling results in high sensitivity to the resolution and quality of the experimental data available, meaning that larger step sizes can be beneficial. In addition, although not typically considered, the experimentally reported energies $E$ also have uncertainties. As such, the barrier distribution must be computed with several different choices of intervals to empirically determine if the result is stable. 

A potentially more numerically robust strategy is to fit a smooth function to the data and take derivatives of the fitted function \citep{Scamps2018}, or to avoid taking second derivatives entirely in favor of computing moments of the distribution function as is done in both \cite{Scamps2018, Rehm2016}. Building on the former advance, \cite{Godbey2024} recently proposed a novel technique to fit Gaussian process models to experimental data, allowing one to take derivatives of a smooth function so long as the chosen kernel was twice differentiable, while also explicitly encoding an uncertainty estimate of the barrier distribution. The initial work was benchmarked on experimental fusion cross section data and showed strong performance in capturing the primary peak of the distribution functions. At large energies, the experimental uncertainties are large enough that it becomes challenging to determine whether the Gaussian process is representing the structure of the distribution function correctly, considering that the estimated barrier in this energy region via the 3-point difference formula is unreliable. 

To better understand the robustness and limitations of the method, this chapter applies the GP method to a toy problem with a known distribution function. We corrupt the ground truth data to mimic the quality and resolution of the fusion datasets used to draw practical conclusions for future applications of this technique to analyze experimental data. 

\begin{figure}[htbp]
  \centering
  \includegraphics[width=0.9\textwidth]{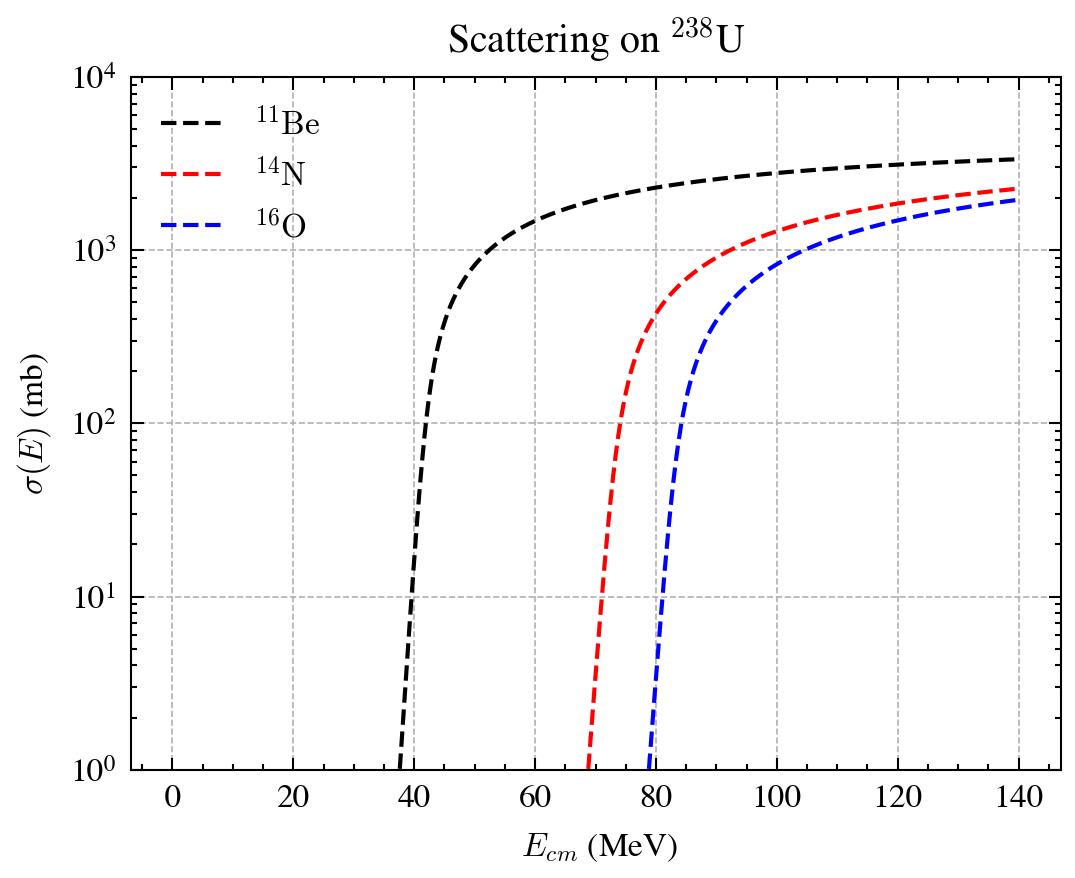}
  \caption{Sample Wong cross sections for different reactions computed with a Woods-Saxon parameterized by $V_0= 70.0~\mathrm{MeV}$, $a=0.48~\mathrm{fm}$, and $r_0=1.25~\mathrm{fm}$.}
  \label{fig:wong-examples}
\end{figure}

\section{Preliminaries}
\subsection{Wong Model of Reaction Cross Sections}
Wong first introduced his formula in 1973 by considering a simple model for the average interaction each partial wave experiences. He took a Woods-Saxon-like potential as an approximation for the mean-field nuclear potential summed with a centrifugal barrier and the Coulomb force. \begin{equation}
    V_l(r) = \frac{Z_1 Z_2 e^2}{r} + \frac{\hbar^2l(l+1)}{2\mu r^2} -V_0[1+\exp((r-R_1-R_2)/a)]^{-1}
\end{equation} With a few empirical approximations and treating the interaction barrier (i.e. the local maximum of the potential through which tunneling takes place) for each partial wave as an inverted harmonic oscillator (also known as a parabolic barrier), he obtained a formula describing the total reaction cross section \begin{equation}
    \sigma(E) = \frac{R_0^2\hbar\omega_0}{2E} \log(1+
    \exp[2\pi (E-E_0)/\hbar \omega_0])\label{eqn:wong}
\end{equation} where $R_0$, $E_0$, and $\hbar \omega_0$ are the location, height, and curvature for the interaction barrier for the s-wave. While in principle, these parameters can be extracted numerically after calibrating the Woods-Saxon (see Figure \ref{fig:wong-examples}), in practice, theorists quickly began to fit to experiment to perform theoretical analysis of experimental results \citep{Vaz1974,  Alexander1975, Eyal1976}. 

\subsection{Gaussian Processes}
Gaussian processes are a powerful Bayesian inference technique that determine a distribution of functions describing a dataset. This allows GPs to a) make predictions and estimate their uncertainty and b) sample numerous functions that could describe the data. Formally, a Gaussian process is defined as a collection of random variables for which any finite subset follows a multivariate normal distribution \citep{RasmussenWilliams2005}. This means that Gaussian processes assume that individual function values $f(x_i)$ of the model that describes the data $f(x)$ are assumed to follow Gaussian distributions.

Consider the 1D task of finding a continuous mean prediction function $\bar f(x)$ with pointwise uncertainties $\delta \bar f(x)$ that describes a collection of observations $(\mathbf{x}, \mathbf{y})$. We could equivalently treat this as a search for a distribution of continuous functions of the form $f(x)$ that describe the observed data. Gaussian processes accomplish this task by claiming the possible values of any function at $x_i$ are given by a Gaussian distribution. A distribution of functions is therefore an infinite collection of individual Gaussian distributions assigned to every real number on the interval of interest. Sampling a point from each individual Gaussian and plotting them would of course look like a structureless scatter plot in the case that each Gaussian is independent of its neighbors. However, by imposing that Gaussians near one another are strongly correlated while being less strongly correlated with Gaussians at $x$ values further away, samples will begin to look more and more smooth and continuous.

In reality, computers do not enable us to store the values of functions evaluated at infinite points from our sample space. Instead, the far more practical solution that contains the equivalent amount of information is to have a ``rule'' to assign outputs when given inputs. Likewise, to make Gaussian processes computationally tractable, we do not have to directly consider a collection of actual functions of closed form. Instead, we can obtain the desired rule exactly due to the immensely powerful marginalization property of Gaussian distributions \citep{Shi2019}. In essence, if one can specify the joint Gaussian distribution of two random vectors (not necessarily of the same dimension) then one has access to the distribution of the individual random vectors. 

\begin{equation}
p(\mathbf{y}_a, \mathbf{y}_b)
\sim
\mathcal{N}\!\Biggl(
\begin{bmatrix}
\mathbf{m}_a \\[3pt]
\mathbf{m}_b 
\end{bmatrix},
\begin{bmatrix}
K_{aa} & K_{ab} \\[3pt]
K_{ba} & K_{bb}
\end{bmatrix}
\Biggr)
\quad\Longrightarrow\quad
p(\mathbf{y}_a)
\sim
\mathcal{N}\!\bigl(\mathbf{m}_a,\,K_{aa}\bigr)
\end{equation} 

where $\mathbf{m}_a$ and $\mathbf{m}_b$ are the means of $\mathbf{y}_a$ and $\mathbf{y}_b$ respectively. This ``decoupling property'' is useful for two reasons.

First, if we consider a function $f(x)$ given by a Gaussian process to be an infinite set of points $\{f(x_i)\}$, we can partition this set into a finite set of all the ``important points'' (i.e. the union of all observed points and any new points we want to query the Gaussian process at) and all other points which form an infinite set. Essentially, we can pick out an arbitrary, finite set of points and our computations are unaffected by the infinite points that we excluded. 

The second utility of this property is that we can again partition the set of ``important points'' into the set of observation points and the points we wish to query at. Bayes' theorem tells us \begin{equation}
    p(\mathbf{y}_{*} |\mathbf{y}) = \frac{p(\mathbf{y}_{*},\mathbf{y} )}{p(\mathbf{y})}
\end{equation}
Each of these follows Gaussian distributions, meaning the conditional distribution is also Gaussian and has the analytic form 
\begin{equation}
\begin{aligned}
p(\mathbf{y}_{*} \mid \mathbf{y})
&= \mathcal{N}\Big(
\mathbf{m}_{*} + K(\mathbf{x}_*,\mathbf{x})K(\mathbf{x},\mathbf{x})^{-1}(\mathbf{y} - \mathbf{m}), \\
&\qquad
K(\mathbf{x}_*,\mathbf{x}_*) - K(\mathbf{x}_*, \mathbf{x})K(\mathbf{x},\mathbf{x})^{-1}K(\mathbf{x},\mathbf{x}_*)
\Big).
\end{aligned}
\label{eq: Noiseless GP}
\end{equation}
where $\mathbf{x}$ and $\mathbf{x}_*$ are sets of observation and query points respectively and $K(*,*)$ is a covariance matrix. For full details of this last step, one can refer to \cite{RasmussenWilliams2005}. So long as the modeler can specify the mean vectors and covariance matrices at the observed and desired points, the GP can be used for practical inference.

To construct the various $K$ matrices, a user can choose a scalar symmetric function $K_{i,j} = k(x_i, x_j) = k(x_j, x_i)$ describing how correlated two points are. Likewise, one can choose the precise form of $\mathbf{m}_i = m(x_i)$ to best reflect their prior knowledge of the underlying function. A default choice of the mean function is simply the zero function. 

This makes Gaussian processes highly flexible, enabling scientists to apply their domain knowledge to effectively model their data by specifying mean and kernel functions. An important strength of GPs is that they can be extended to reflect noise in observed data by assuming the noise is Gaussian and adding a variance to the diagonal terms of the observation covariance matrix. 

\section{Characteristic Experimental Data Available}
\begin{figure}[htbp]
\includegraphics[width=\textwidth]{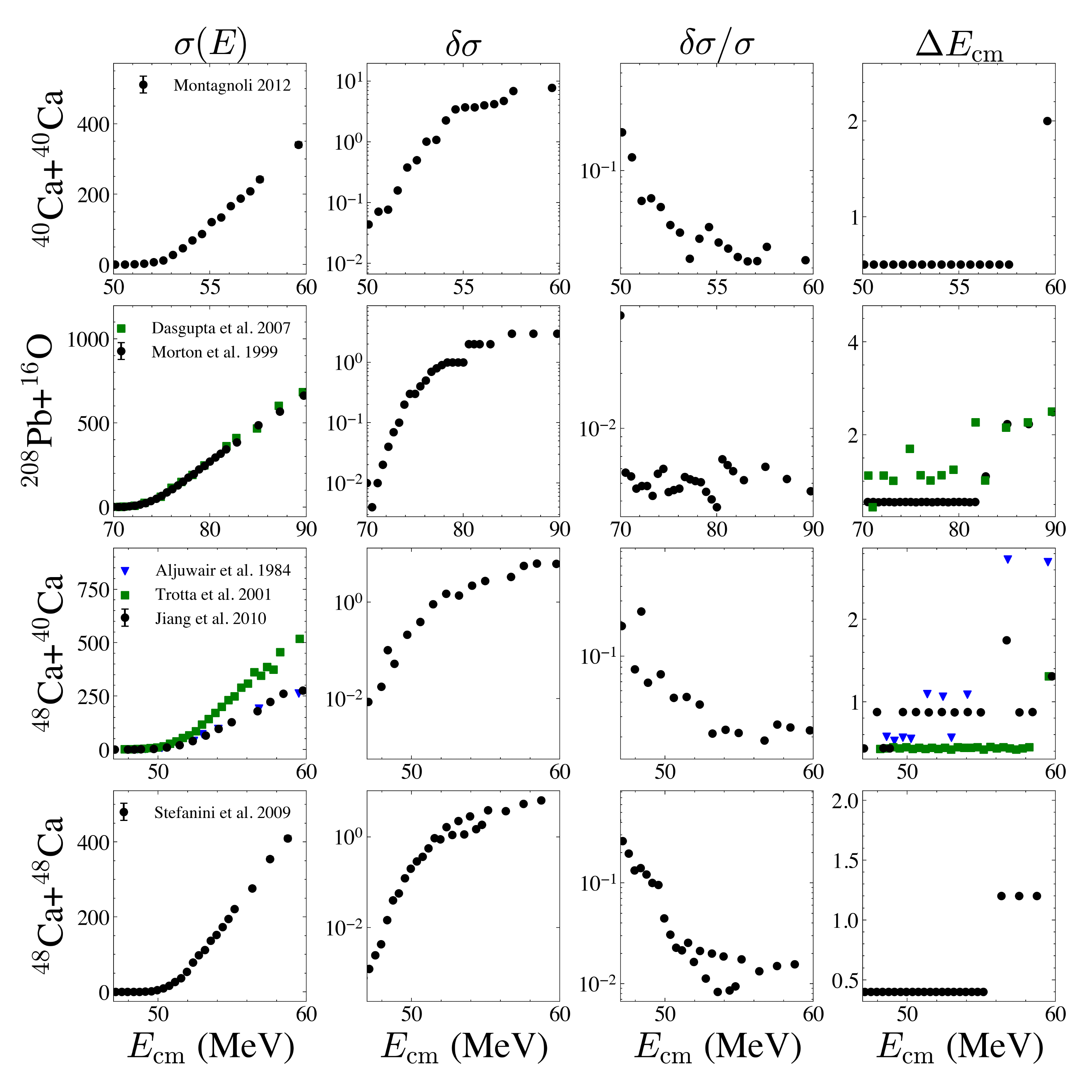}








 \caption{Experimental fusion cross sections for $^{16}$O + $^{208}$Pb \citep{Morton1999, Dasgputa2007}, $^{40}$Ca+$^{40}$Ca \citep{Montagnoli2012}, $^{48}$Ca+$^{40}$Ca \citep{Aljuwair1984, Jiang2010, Trotta2001}, and $^{48}$Ca+$^{48}$Ca \citep{Stefanini2009}. Quoted uncertainties appear in the second and third column when available.}
 \label{fig:exp_data}
\end{figure}

Figure \ref{fig:exp_data} demonstrates some basic properties of fusion cross section data for medium and heavy ion collisions. Of interest were the resolution, error scale, and irregularity of the data near the barrier. In general, measurements near the primary fusion barrier had resolutions between 0.4 MeV $\leq \Delta E \leq$ 1.0 MeV. Rather than examining the range of quoted experimental uncertainties of the cross section, it is more illustrative to inspect the ratio of uncertainties to the reported value of the cross section $\eta \equiv \delta \sigma/\sigma$. Given that cross sections can vary orders of magnitude across energies, this ratio is a more robust metric of the noise present in each dataset. Empirically, we found different datasets had relatively constant $\eta$ across all points. Across all datasets, the noise level ranged between $\eta=10^{-2}$ (high precision) to $\eta=10^{-1}$ (lower precision).

To quantify the irregularity of the data spacing, we computed the Coefficient of Variation (CV, denoted by $\kappa$ in plots) of step sizes $\Delta E$. CVs are used to quantify the relative spread of samples in a distribution and are defined by dividing the standard deviation of a dataset by its mean. Many datasets had approximately constant step sizes near the barrier, but generally CVs ranged from 0 $\leq \kappa\leq$ 0.7. 

Exact computations of all relevant statistics can be found in a Jupyter notebook on the accompanying repository \citep{Philip2026}.

\section{Methods}
We use the Wong formula given in Equation \ref{eqn:wong} as a toy model to test the performance of the Gaussian process regression technique. We simulate datasets with similar statistics to the real experiments in Fig. \ref{fig:exp_data} to investigate how robust the GPR formalism is in capturing data with different resolution $\Delta E$, noise levels $\eta$, and irregularity $\kappa$.

\begin{figure}[htbp]
  \centering
  \includegraphics[width=0.8\textwidth]{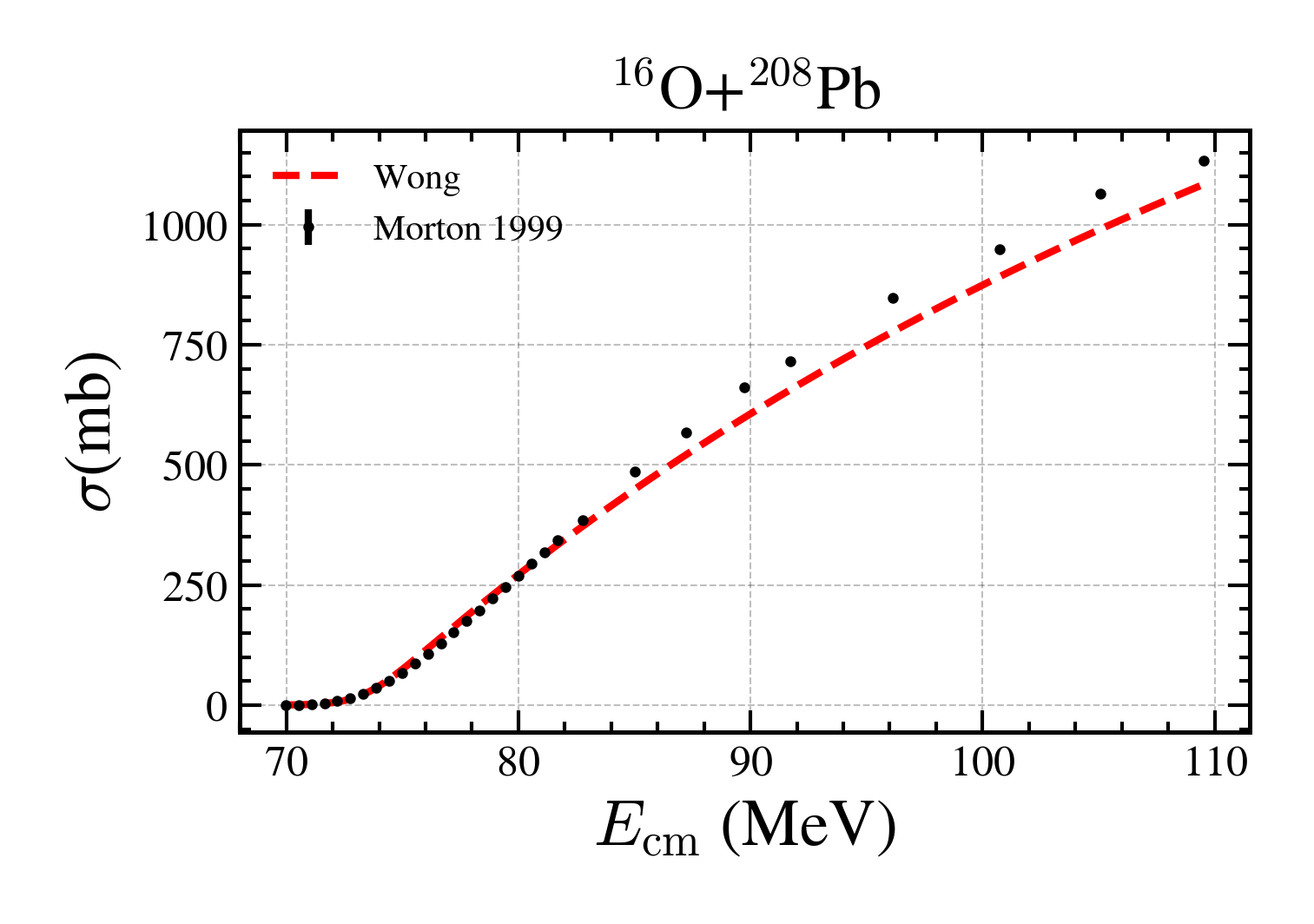}
  \caption{Chi‐squared fit of the Wong formula in Eq.~\ref{eqn:wong} at parameters $E_0=73.38 $ MeV, $R_0 =10.23$ fm, and $\hbar\omega=4.84$ MeV.}
  \label{fig:wong-chi-squared}
\end{figure}

To obtain ground truth for the simulated data, we used $\chi^2$ regression to fit the data in \cite{Morton1999}, given that it generally had the highest quality data with high resolution and regular measurements near the barrier as well as a low ratio of experimental uncertainty to measurement. Figure \ref{fig:wong-chi-squared} shows the corresponding curve that minimized the $\chi^2$ score over the whole dataset which is given by the parameters $E_0=73.3846 $ MeV, $R_0 =10.2260$ fm, and $\hbar\omega=4.8410$ MeV.

To extract barrier distributions in all future experiments, we use finite differences on a high-resolution grid upon which we query the Gaussian process post-training. As in the original study by \cite{Godbey2024}, we empirically found this to be relatively numerically stable under different grid resolutions. Like the original study, we use a zero-mean Gaussian process with Radial Basis Function kernel given by \begin{equation}
    K_{ij} = k_{RBF}(x_i, x_j) = C\exp(-\frac{(x_i-x_j)^2}{2l^2})
\end{equation} 
This kernel was originally chosen as it is infinitely differentiable--although other kernel structures were experimented with, they yielded little improvement. The parameters $C, l$ are adjusted to maximize likelihood during training.

An optional White Noise kernel was also used in the original study \citep{Godbey2024} with a learnable shared homoscedastic term $\sigma_n^2$
\begin{equation}
    k(x_i, x_j) = k_{RBF}(x_i, x_j) + \sigma_n^2 \delta_{ij},
    \label{eq: whitenoisekernel}
\end{equation}

Alternatively, adding a small, non-negative number to the diagonal of the covariance matrix is generally done to increase numerical stability during inversion of the covariance matrix. One interpretation of this regularization is that adding the variance of the estimated noise to the diagonal of the covariance matrix can explicitly encode prior information about the reliability of observed data, by assuming the errors follow a Gaussian distribution. Thus, an alternate strategy is to replace $K(X,X)$ with 
\begin{equation}
    K_y = k_{RBF}(x_i, x_j) +\delta\sigma_{i}^2 \delta_{ij},
    \label{eq:diagonal errors}
\end{equation}
when experimental errors $ \delta\sigma_i^2$ are known.




\subsection{Additive vs. Multiplicative Error Model}
\begin{figure}[htbp]
   \centering
   \includegraphics[width=0.9\textwidth]{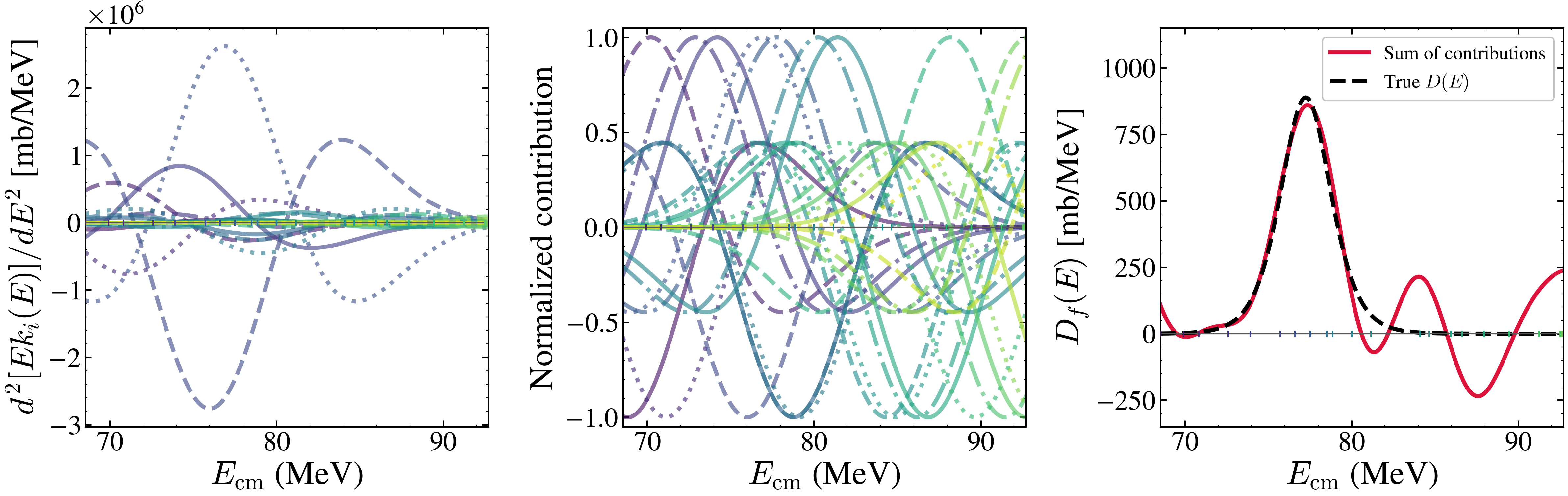}

   \caption{ Decomposition of a mean prediction for the barrier distribution into a linear combination of kernel functions centered on training data locations. (Left) Each component term multiplied by its corresponding weight. (Central) Normalized kernel contributions to showcase the variety of curves summed. (Right) The corresponding mean prediction. Note that there are huge cancellations of contributions. 
   }
   \label{fig:gp-decomposition}
\end{figure}

The closed-form expression in Eq. \ref{eq: Noiseless GP} for the posterior over values $\mathbf{y_*}$ at inference locations $\mathbf{x_*}$ is applicable when observations $(\mathbf{x}, \mathbf{y})$ are noiseless. However, if one instead assumes that the observed data follows an additive error structure i.e. that 
\begin{equation}
\begin{aligned}
    y_i &= f(x_i) + \epsilon_i, \\
    \epsilon_i &\sim \mathcal{N}(0,\delta y_i^2).
\end{aligned}
\end{equation}

Utilizing the kernel given by Eq. \ref{eq:diagonal errors}, one can instead attempt to infer the noiseless model from noisy observations via 

\begin{equation}
\begin{aligned}
p(\mathbf{f}_{*} \mid \mathbf{y})
&\sim \mathcal{N}\Big(
\mathbf{m}_{*} + K(\mathbf{x}_*,\mathbf{x})K_y^{-1}(\mathbf{y} - \mathbf{m}), \\
&\qquad
K(\mathbf{x}_*,\mathbf{x}_*) - K(\mathbf{x}_*, \mathbf{x})K_y^{-1}K(\mathbf{x},\mathbf{x}_*)
\Big).
\end{aligned}
\label{eq: Noisy GP}
\end{equation}

If $\mathbf{m_*}, \mathbf{m}$ are chosen to be zero, the mean prediction reduces to $K(\mathbf{x}_*,\mathbf{x})K_y^{-1}\mathbf{y}$. The mean prediction can be written in continuous form as 
\begin{equation}
    \bar f(x) = \sum _i^n k(x,\mathbf{x}_{i})(K_y^{-1}\mathbf{y})_i
\end{equation}
This effectively decomposes mean predictions into a sum of kernels centered on data points. In the case of $k_{RBF}$, the mean prediction is a sum of squared exponentials centered on the locations of training points. Therefore, if we seek to fit the mean to the underlying excitation function $\sigma(E)$ by assuming experimental measurements $(\mathbf{E}, \mathbf{y_{\sigma}}, \mathbf{\delta y_{\sigma}}) \in (\mathbb{R}^{n}, \mathbb{R}^{n}, \mathbb{R}^{n})$ arise from an additive error model, we can also analytically examine the resulting barrier distribution: 
\begin{equation}
    \frac{d^2}{dE^2}[Ef_\sigma(E)] = \sum _i^n  (K_y^{-1}\mathbf{y_{\sigma}})_i \frac{d^2}{dE^2}[E k(E,\mathbf{E}_i)]
\end{equation}

Fig. \ref{fig:gp-decomposition} visualizes what each of these individual component functions look like for an example train to noisy measurements from a Wong formula curve. The left panel visualizes what each continuous component looks like while the central panel normalizes each component to visualize the curve structure. The right panel shows the mean prediction obtained as a sum of all curves shown in the left panel. Immediately, it becomes apparent that the trained Gaussian process obtains its mean prediction through massive cancellations of component functions. Across many scenarios, we observed this same feature. Further investigations indicate that the combination of multiscale data and irregular spacing leads to an ill-conditioned covariance matrix $K_y$ such that elements of $(K_y^{-1}\mathbf{y})$ can span several orders of magnitude for a given train. 

To circumvent this issue, we instead train all models in log space. To determine corresponding variances, we apply the delta-method \cite{casella2024}. The new quantities are given by
\begin{equation*}
\begin{aligned} 
\mathbf{z_{\sigma}} = \log (\mathbf{y_{\sigma }}) \\
    \delta\mathbf{z_{\sigma}} \approx \delta \mathbf{y_{\sigma}} / \mathbf{y_{\sigma}}
    \end{aligned}
\end{equation*} 
with both equations defined component-wise. Smooth samples from the GP are subsequently exponentiated back into linear space. Employing an additive error model in log space is equivalent to utilizing a multiplicative error model in linear space: one where uncertainties scale with the size of the observations. 

\begin{equation}
\begin{aligned}
    z_i &= \log y_i,
    \qquad
    g(x_i)=\log f(x_i),
    \qquad
    \delta z_i \approx \frac{\delta y_i}{y_i}, \\
    z_i &= g(x_i)+\eta_i,
    \qquad
    \eta_i \sim \mathcal{N}(0,\delta z_i^2)
    \quad \Longleftrightarrow \quad
    y_i = f(x_i)\xi_i,
    \qquad
    \xi_i \sim \operatorname{LogNormal}(0,\delta z_i^2).
\end{aligned}
\end{equation}

As seen from the 3rd column of Fig. \ref{fig:exp_data}, this is a reasonable statistical model for the fusion data we are interested in. We empirically found this strategy stabilized results across a host of realistic experimental conditions and employ it for the rest of the thesis. 

Although the log-space training strategy does improve stability it does not remove the numerical aliasing of false peaks at higher energies as the following results demonstrate. The component functions are oscillatory as shown by the central panel of Fig. \ref{fig:gp-decomposition}, and the remnants of the large cancellations of components typically also exhibit oscillatory behavior. 

\section{Results}
 \begin{figure}[htbp]
   \centering
\includegraphics[width=0.9\textwidth]{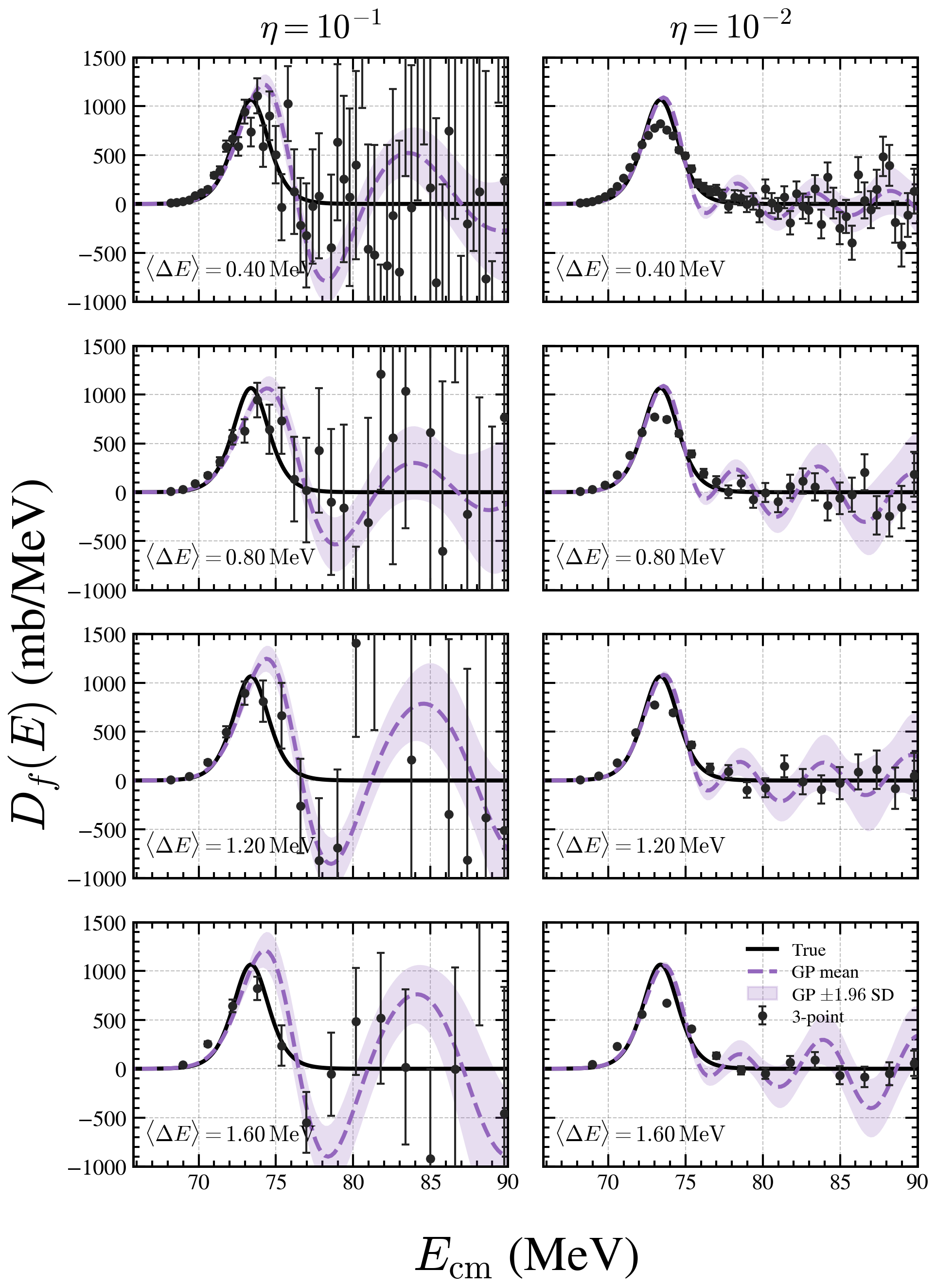}

  \caption{Left and right columns examine how extracted distribution functions differ given different data resolutions near the barrier for two realistic bounds on estimated experimental uncertainties. The spacing on the 3-point formula was chosen such that the average spacing was closest to 2.5 MeV. }
   \label{fig:gp-data-resolution}
 \end{figure}
 
\subsection{Data Resolution}

Given that GPs stopped capturing the barrier location between $\eta=10^{-2}$ and $\eta=10^{-1}$ in the preliminary tests, we trained GPs under these two distinct noise structures while varying the resolution of the training data in realistic ranges reflecting the available experimental data. 

Figure \ref{fig:gp-data-resolution} shows results for both noise models: each column shows the effect of varying the resolution at a fixed noise level. At high noise levels ($\eta=10^{-1}$), the model is not able to capture the location of the barrier even under high resolution data, and both the width and height of the predicted barrier decay in quality as the training data resolution is lowered. Moreover, the GP predicts secondary peaks that do not reflect the true distribution well and are smoothed away only by lowering training resolution. This is a commonly observed characteristic of radial basis kernels.
With the lower noise data ($\eta=10^{-2}$) by contrast, the GPs capture the height, location, and width of the primary peak robustly even as the data resolution is reduced considerably. Although these GPs also predict some secondary periodic structure at higher energies, they are much better constrained and have sufficient spreads of predictions that convey a lack of certainty in higher energy structures until the training resolution is made coarser. Thus, we conclude that noise in the observed data is more important than sparsity of data when fitting GPs both in capturing primary structure of the barrier and in representing the distribution structure and uncertainty at higher energies.

\begin{figure}[htbp]
   \centering
\includegraphics[width=0.9\textwidth]{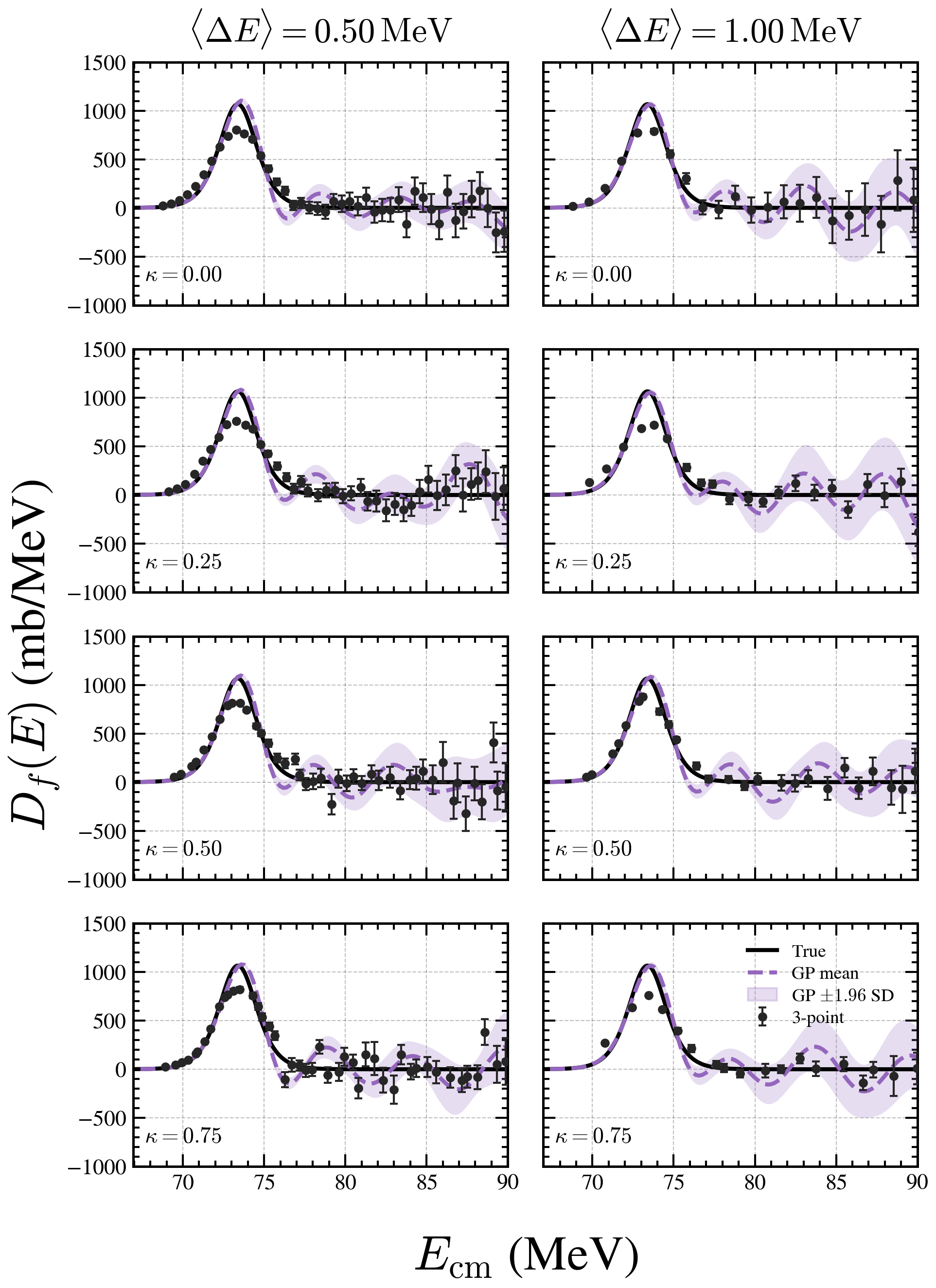}

   \caption{Effect of irregular data on extracted distribution functions. Note $\eta=10^{-2}$ for all plots to maximally isolate the effect of the irregularity of the data. The spacing on the 3-point formula was chosen such that the average spacing was closest to 2.5 MeV.}
   \label{fig:gp-data-irregularity}
 \end{figure}
 
\subsection{Data Irregularity}
Given the results of the previous experiment and the considerably better performance in capturing the main features of the barrier, we fix the observation noise level at $\eta=10^{-2}$ for this experiment. Given the range of step sizes and the irregularity of these steps in the experimental fusion cross sections, we created datasets with 4 different CV in step sizes for an average step size of $\Delta E=0.50$ MeV and $\Delta E=1.0$ MeV. So a CV = 0.0 with mean step size $\Delta E=0.50$ MeV corresponds to a perfectly regular grid where all measurements are taken every 0.50 MeV, whereas higher CVs correspond to more variations in step sizes, while still being centered at 0.5 MeV. Figure \ref{fig:gp-data-irregularity} highlights that the GPR formalism is actually very robust under irregular data. Although results are generally more representative of the true distribution when the average step size is 0.5 MeV, all GPs trained capture the height, width, and location of the single barrier remarkably well regardless of data resolution or irregularity. The coarser data does result in some aliasing of secondary structure at higher energies as we observed in the previous experiment, but overall the high-energy distribution uncertainty and structure are both well estimated for the finer, more regular grids. 

\subsection{Inferring Unknown Relative Uncertainty}
\begin{figure}[htbp]
   \centering
\includegraphics[width=0.9\textwidth]{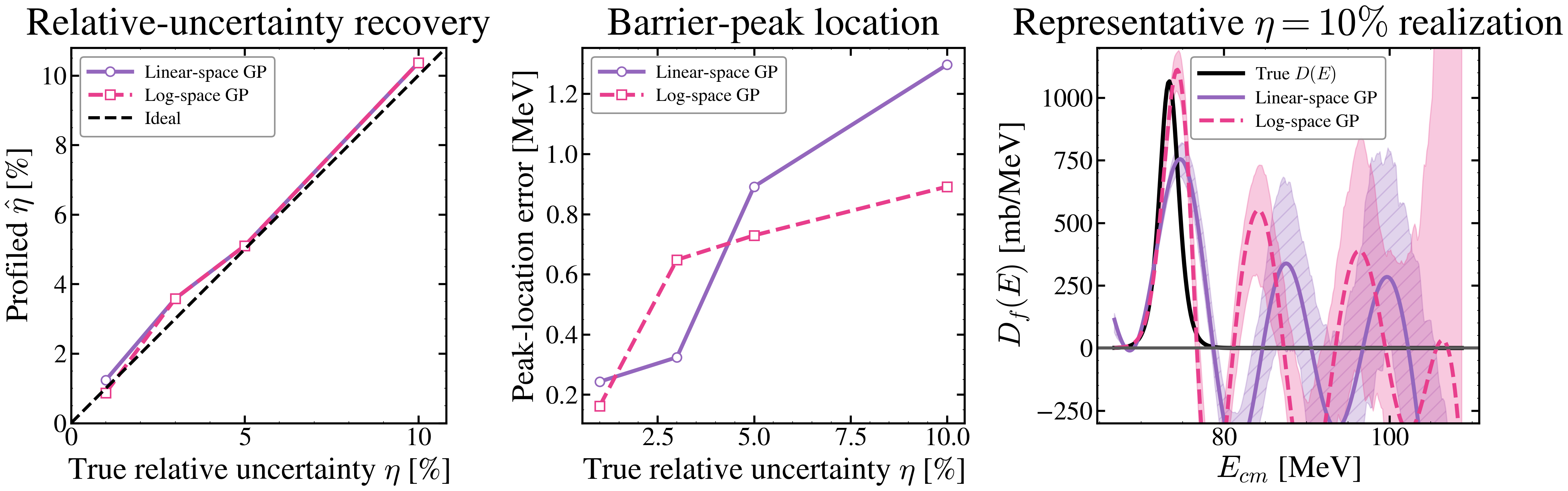}

   \caption{(Left) Comparison of how well GPs with White Noise kernels can infer the true relative error via marginal likelihood maximization in linear and logarithmic space. (Center) Ability of each method to recover the location of the primary peak. (Right) Representative learned barrier distributions with each method. }
   \label{fig:gp-inferred-eta}
 \end{figure}

In situations where no quoted uncertainty was available, one may turn to a White Noise kernel and optimize a global variance over all measurements during fitting to maximize the log likelihood. This is the intended purpose of a White Noise kernel--to model unknown homoscedastic noise. This becomes an even more effective strategy when training in log space given that the relative error is often approximately constant. However, using the White Noise as part of the chosen kernel $k(*,*)$ to sample curves (i.e. applying Eq. \ref{eq: Noiseless GP} with the kernel from Eq. \ref{eq: whitenoisekernel}) will result in samples of $\mathbf{y_*}$ being drawn and differentiated. Recall that the purpose of utilizing Bayesian inference in this situation was to infer the possible smooth, latent functions that could have plausibly generated the noisy observed data. Therefore, sampling $\mathbf{y_*}$ and differentiating creates numerically unstable derivatives because differentiation amplifies the effect of uncorrelated noise corrupting the smooth latent function. 

Therefore, we utilized an altered strategy. In this method, we perform a standard GP regression utilizing a White Noise kernel. The optimization algorithm determines the most likely global variance describing the data by maximizing the marginal likelihood. At inference time, we remove the White Noise kernel and instead use the learned global variance to construct $K_y$. Consequently, we sample and differentiate $\mathbf{f_*}$ rather than $\mathbf{y_*}$. Fig. \ref{fig:gp-inferred-eta} illustrates the relative performance of this strategy in log space and linear space alongside a sample Wong formula. While both methods can reasonably recover the original relative uncertainty, the log-space training strategy is generally able to recover the primary barrier with greater fidelity. The tradeoff is that the log-space GP is less reliable at higher energies than the GP trained in linear space. We utilize this method in Ch. 4 to analyze experimental measurements where quoted uncertainties were unavailable.

\section{Conclusions}

The main conclusions relevant to modeling these barrier distributions with Gaussian processes were that \begin{enumerate}
    \item Being able to explicitly specify uncertainty in observed data as well as the freedom to choose a non-zero mean function as a prior is a key strength of the Gaussian process. The BNNs we test in the next chapter have neither of these properties.
    \item More than any other factor, considerable noise in the observed data affects a GP's ability to accurately reproduce the barrier. 
    \item If the data is relatively noise free, the Gaussian process is extremely robust to variations in data resolution and irregularity of the grid.
    \item Radial Basis Functions introduce aliasing of secondary structure at energies higher than the initial barrier's location. These effects are minimized when the data is high resolution, regular, and noise-free but are still present. It is difficult to trust the predictions of higher energy structure unless the fusion datasets are remarkably high quality. 
    \item Training in logarithmic space helps stabilize GP results as it corresponds to employing a multiplicative error model that is consistent with the excitation function data we are analyzing.  
\end{enumerate}

\clearpage
\chapter{Bayesian Neural Networks for Sampling Barrier Distributions}\label{c:chap3}
\chaptermark{}

\section*{Abstract}
In Chapter 2, we demonstrated situations where Gaussian processes struggle to represent features of the true barrier and quantify its uncertainties. One approach to improve performance is to experiment with different covariance kernels and combine simpler kernels into more complex ones. However, we are still limited by the implicit assumption that model predictions and observational errors are both given by normal distributions. 

In this Chapter, we explore interpretable models for our data using interpretable Bayesian Neural Networks (BNNs) with flexible parameter posteriors. We utilize the AutoBNN software tool released by Google to explore BNN architectures that are shown to correspond to sophisticated Gaussian process kernels in the limit of large layer widths. We describe the computational infrastructure required to obtain and sample the posteriors and benchmark results on several of the same toy datasets used in Chapter 2. We find that certain AutoBNN architectures are more robust in representing the barrier at higher energies, and that others share many characteristics with Gaussian processes.

\section{Background}
To understand how AutoBNN is constructed and trained, we first investigate the more general problem of statistical modeling before discussing the unique challenges that models with many free parameters face. The following content is my own presentation of many ideas reviewed in \cite{Jospin2022, Blei2017, casella2024} and these excellent resources should be consulted for greater detail and other interesting research. 

\subsection{Bayesian Inference for Parameters}
As scientific modelers, we often calibrate a proposed parametric model $f_{\boldsymbol{\theta}}(\mathbf{x})$ to measured data $D$ by adjusting internal parameters $\boldsymbol{\theta}$ to reproduce the data. If these measurements have some element of uncertainty or error, we can reflect this in our model by treating the parameters $\boldsymbol{\theta}$ as arising from a probability distribution conditioned on observed data denoted by $p(\boldsymbol{\theta} | D)$. We can make predictions using parameters that are \textit{most likely} to explain the data, but importantly, we can also quote uncertainties by considering predictions arising from less likely parameter choices. In other words, we could sample many different parameters from $p(\boldsymbol{\theta} |D)$ and consider the distribution of predictions made by the induced distribution of functions $f_{\boldsymbol{\theta}} \sim p_f(f_{\boldsymbol{\theta}}|D)$. This probabilistic strategy does not mean our model is inherently non-deterministic; rather, it offers an elegant way to book-keep sources of uncertainty and explicitly acknowledge limitations of our models and assumptions.

The task of obtaining a calibrated model must therefore come down to determining $p(\boldsymbol\theta |D)$ for a given family of parameterized models $f_{\boldsymbol{\theta}}$. Bayes' Theorem tells us the posterior distribution $p(\boldsymbol{\theta}|D)$ is defined as 
\begin{equation}
    p(\boldsymbol\theta|D) = \frac{p(D|\boldsymbol{\theta})p(\boldsymbol{\theta})}{p(D)} = \frac{p(D|\mathbf{\boldsymbol\theta})p(\boldsymbol{\theta})}{\int p(D|\boldsymbol{\theta}')p(\mathbf{\boldsymbol\theta} ') d\boldsymbol{\theta '}}
    \label{eq:parametric-bayes}
\end{equation}
where the \textit{likelihood} $p(D|\boldsymbol\theta)$ describes how suitable each choice of parameters is for describing the data and the \textit{prior} reflects our beliefs about what the parameter distribution $p(\boldsymbol\theta)$ could look like before we ever see data. Meanwhile, the \textit{evidence} $p(D)$ can be re-expressed as an integral over all possible choices of parameters. Because the evidence marginalizes (i.e. integrates out unwanted degrees of freedom) over $\boldsymbol\theta$, it is solely dependent on the data $D$. Treating the evidence as essentially a normalization constant is key in scenarios where the marginalization cannot be explicitly performed as we will discuss soon. 

\subsection{Example: Maximizing the Posterior}
\label{subsect:example-maximizing-posterior}
Let's consider an explicit example. Suppose we have $n$ observations of the form $(x_i, y_i) \in \mathbb{R}^2$ composing our dataset $D$ where there is additive error only on $\mathbf{y}=(y_1,...,y_n)$ given by \begin{equation}
\begin{aligned}
       y_i =f(x_i) +\epsilon_i, \\
       \epsilon_i \sim \mathcal N(0,\sigma^2)
\end{aligned}
\end{equation}
Note that the noise is uncorrelated among measurements. Moreover, assume for now that we choose a relatively uninformative prior given by a uniform distribution in some reasonably large but finite range $p(\boldsymbol\theta) = \mathrm{Uniform}([\boldsymbol{\theta_1}, \boldsymbol{\theta_2}])$. 

These approximations make it straightforward to find a $\boldsymbol{\hat\theta}=\mathrm{argmax}_{\boldsymbol{\theta}}[p(\boldsymbol{\theta}|D)]$. As stated, the evidence is effectively a normalization constant because the data is fixed during our search for $\boldsymbol{\hat\theta}$. The uniform prior $p(\boldsymbol\theta)$ is a positive constant function in $[\boldsymbol{\theta_1}, \boldsymbol{\theta_2}]$ and 0 everywhere else. Therefore, 
\begin{equation*}
    \boldsymbol{\hat\theta}=\mathrm{argmax}_{\boldsymbol{\theta}}[p(\boldsymbol{\theta}|D)] = \mathrm{argmax}_{\boldsymbol{\theta}}[p(D|\boldsymbol{\theta})]
\end{equation*}
according to Eq.~\ref{eq:parametric-bayes}. 

Under our simple additive error model with independent and identically distributed noise (IID), the likelihood factors\footnote{Perhaps the most useful approximation for analyzing multivariate distributions is the assumption of independence. However, this can be a poor assumption, especially in experimental contexts where systematic measurement errors are correlated due to a miscalibrated detector etc.} as 
\begin{equation}\begin{aligned}
        p(D|\boldsymbol{\theta}) = \prod_{i=1}^n p(y_i |\boldsymbol{\theta}) = \prod_{i=1}^n \frac{1}{\sqrt{2\pi\sigma^2}}\exp[-\frac{(y_i-f_{\boldsymbol{\theta}}(x_i))^2}{2\sigma^2}] \\
        =\frac{1}{(2\pi\sigma^2)^{n/2}}\exp[-\frac{1}{2\sigma^2}\sum_{i=1}^{n}(y_i-f_{\boldsymbol{\theta}}(x_i))^2]
\end{aligned}
\end{equation}
Thus, we finally see that the optimal parameters are given by
\begin{equation}
    \begin{aligned} \boldsymbol{\hat\theta}= \mathrm{argmax}_{\boldsymbol{\theta}}[-\sum_{i=1}^{n}(y_i-f_{\boldsymbol{\theta}}(x_i))^2]\\
= \mathrm{argmin}_{\boldsymbol{\theta}}\sum_{i=1}^{n}(y_i-f_{\boldsymbol{\theta}}(x_i))^2    = \mathrm{argmin}_{\boldsymbol{\theta}}(-\log p(\boldsymbol{\theta} | D))
    \end{aligned}
\end{equation}

If you have worked with regression at some point (whether it be training a neural network or fitting a linear curve to data) you recognize this result: to find the ``best'' parameters describing the data, one should find the parameters that will minimize the $\ell^2$ norm of the residual. In this case, $\boldsymbol{\hat \theta}$ is known as a \textit{Maximum Likelihood Estimator}. However, the reason for demonstrating this somewhat contrived example is to offer some more insight into what is really meant by ``best''. In this case, we mean the parameters that maximize the given posterior. Or equivalently, the parameters that minimize the negative log posterior. Note that $\mathrm{argmin}_{\boldsymbol{\theta}}(-\log p(\boldsymbol{\theta} | D)) = \mathrm{argmax}_{\boldsymbol{\theta}}(p(\boldsymbol{\theta} | D))$ always, regardless of the chosen error structure or model family $f_{\boldsymbol{\theta}}$ we had chosen since $-\log(x)$ is a monotonically decreasing function for $x>0$. Expressing the optimization task in this form allows us to rely on intuition we have for thinking about standard regression as minimization of a cost function. 

Consider each modeling choice we made and how $-\log p(\boldsymbol{\theta} | D)$ and therefore $\boldsymbol{\hat \theta}$ could change as we relax, tighten, or otherwise deform our statistical framework. For example, if the noise scale $\sigma_i$ was non-constant, then $ \boldsymbol{\hat\theta}= \mathrm{argmin}_{\boldsymbol{\theta}}\sum_{i=1}^{n}\frac{(y_i-f_{\boldsymbol{\theta}}(x_i))^2}{\sigma_i^2}$ which corresponds to $\chi^2$ minimization. If the prior was instead chosen to favor more central values of $\boldsymbol{\theta}$ by choosing $p(\boldsymbol{\theta}) = \mathcal N(\boldsymbol{0}, \Sigma_p^2)$ with $\Sigma_p$ diagonal, then the negative log posterior has an additional term $\propto \sum_{i=1}^{N_p} \theta_i^2$ often called a \textit{regularizer} in the machine learning community. And if $\boldsymbol{\epsilon} \sim\mathcal N(\mathbf{0}, \Sigma_{\epsilon})$ for arbitrary $\Sigma_{\epsilon}$ (i.e. noise is allowed to be correlated), one obtains the generalized result that $-\log p(\boldsymbol{\theta} | D)\propto (\mathbf{y}-\mathbf{f}_{\boldsymbol{\theta}})^T\Sigma_{\epsilon}^{-1}(\mathbf{y}-\mathbf{f}_{\boldsymbol{\theta}}) + \log|\Sigma_{\epsilon}|$. As such, there is an important relationship between posteriors and cost functions. Note that we analyzed the 1D case for clarity, but that these conclusions extend to arbitrary maps $f_{\boldsymbol{\theta}}:\mathbb{R}^{d_x}\to \mathbb{R}^{d_y}$

\begin{figure}
    \centering
    \includegraphics[width=0.9\linewidth]{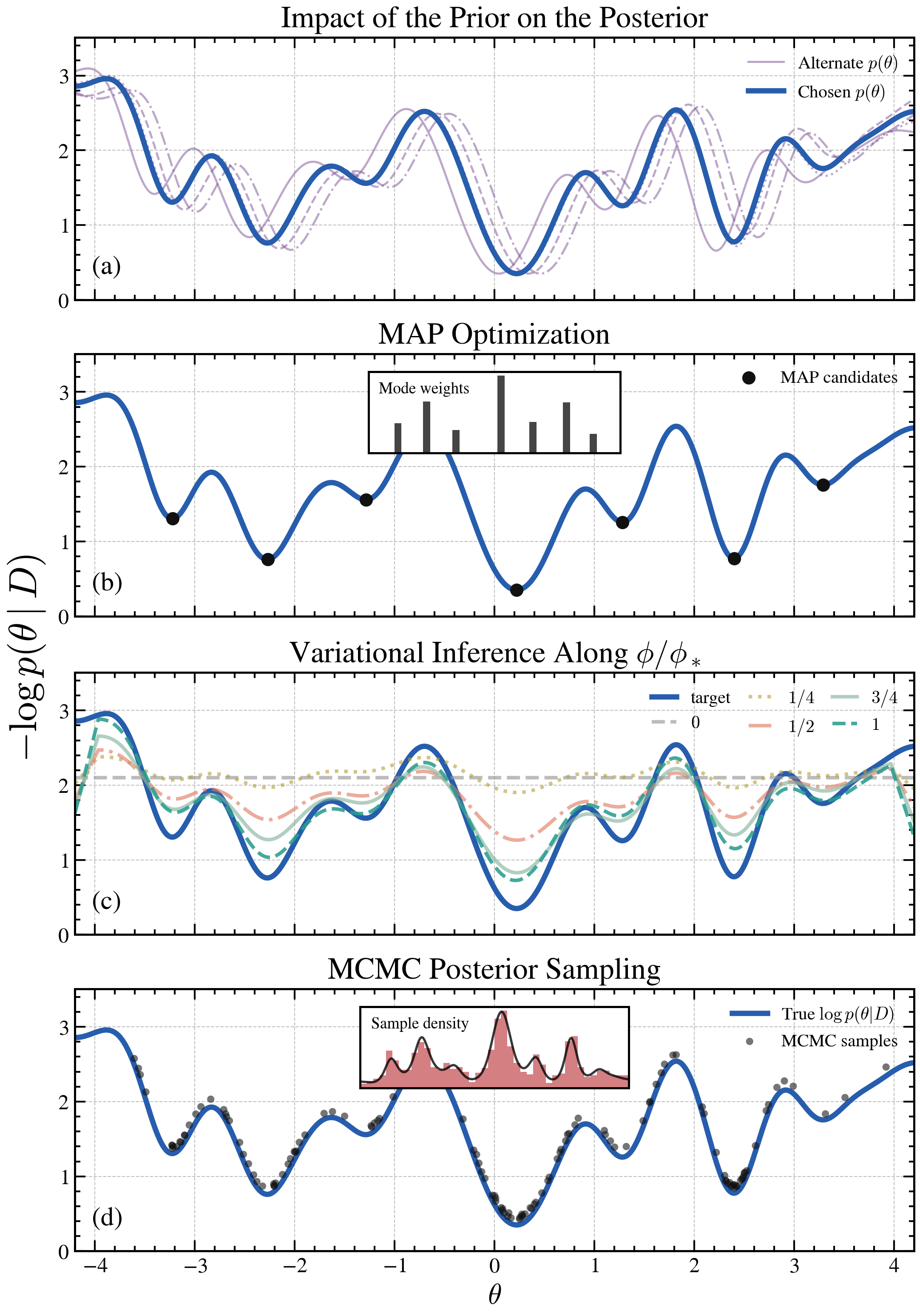}
    \caption{Strategies for training BNNs in analogy to cost functions. See text for details.}
    \label{fig:posterior-visualizatoin}
\end{figure}

\subsection{Posteriors for High-Dimensional Model Parameters}
Till now, we have made almost no assumptions about what $f_{\boldsymbol{\theta}}(\mathbf{x})$ looks like. But suppose $f_{\boldsymbol{\theta}}(\mathbf{x})$ represents a standard neural network\footnote{This thesis has a strong focus on neural networks simply because AutoBNN models are neural networks. In reality, most of the discussion is applicable to any sort of model, especially those with a huge number of free parameters. Moreover, much of the deep learning community's nomenclature is often more familiar which will hopefully aid in connecting together concepts} with anywhere from $\mathcal O(10^{2+})$ individual parameters which collectively compose $\boldsymbol{\theta}$. Indeed the expressivity of neural networks is intimately tied to the huge number of internal parameters in tandem with relatively unobtrusive\footnote{Unobtrusive here means we are using very little of the domain knowledge we may have about the system we are modeling. For example, incorporating known symmetries into a model can massively reduce the number of free parameters to optimize and significantly improve extrapolation power. However, the tradeoff is generally a loss of expressivity.} compositions of linear and nonlinear transformations. Regression or \textit{training} is generally posed as an optimization problem of finding the $\boldsymbol{\hat \theta}$ that minimizes some cost function. Translated, a trained neural network consists of the assumed parametric model or \textit{architecture} $f_{\boldsymbol{\theta}}(\mathbf{x})$ alongside the choice of parameters or \textit{weights} $\boldsymbol{\hat \theta}$ that minimizes some negative log posterior $[-\log p(\boldsymbol{\theta} | D)]$ or \textit{cost function}.

The cost function can be modified by adding regularizer terms to penalize certain behaviors. Noting that
\begin{equation}
    -\log p(\boldsymbol{\theta} | D) = -\log p(D| \boldsymbol{\theta}) -\log p(\boldsymbol{\theta}) + C
\end{equation}
it is clear that adding such regularizers is equivalent to imposing a prior. Fig.~\ref{fig:posterior-visualizatoin}(a) offers a schematic picture in the case that the model family is parameterized by a single number $\theta$. Different priors will result in different negative log posteriors for fixed data\footnote{I keep stressing that the data is fixed to distinguish from cases where new data is regularly becoming available and posteriors update. In these scenarios, an existing posterior can become the new prior each time new data becomes available.} $D$. In the language of machine learning, different choices of cost functions with various regularizers can induce different loss landscapes that can make finding a global minimum (often with some gradient-based method) easier or harder. Regularization can be very effective, but often there is a limit as to how much cost function engineering will improve the loss landscape if a problem is ill-posed.

We've outlined how training large models corresponds to finding a choice of parameters $\boldsymbol{\hat \theta}$ that maximize the value of the posterior $p(\boldsymbol{\theta} | D)$. Doing this discards a lot of information the posterior may contain. For example, the loss landscape may have a very shallow curvature near the global minimum $\boldsymbol{\hat \theta}$, implying that we could choose many different values of $\boldsymbol{ \theta}$ in a neighborhood around $\boldsymbol{\hat \theta}$ and obtain very similar model predictions. What about if there exists another minima $\boldsymbol{\hat \theta'}$  such that $-\log p(\boldsymbol{\hat \theta} | D)$ is slightly lower than $-\log p(\boldsymbol{\hat \theta'} | D)$ but $\boldsymbol{\hat \theta}, \boldsymbol{\hat \theta'}$ are not near one another? Indeed the question of which of $\boldsymbol{\hat \theta}, \boldsymbol{\hat \theta'}$ is the global minima may depend on the chosen prior. In such a situation where the posterior is \textit{multi-modal}, relatively different parameterizations of the same model $f_{\boldsymbol{\theta }}$ may do a similarly good job of describing the data. 

In principle, we know the solution to this dilemma: sample parameters (and therefore different functions) from the full posterior $p(\boldsymbol{\theta} | D)$. \textit{Bayesian Neural Networks} (BNNs) do exactly this: they treat the internal weights of the network as having been drawn from a distribution. By sampling different weights and performing inference on the same input sample, one can achieve a spread of predictions that can capture the BNN's uncertainty. Choices of $\boldsymbol{\theta}$ for which the posterior is larger will get sampled more often, helping us form a central prediction. Meanwhile, the ability to sample the posterior offers additional insight into which areas of the domain our model has the greatest uncertainty. Such values of $\mathbf{x}$ are characterized by point-wise distributions $\{f_{\boldsymbol{\theta_i}}(\mathbf{x})\}_{\forall i}$ that are very wide or multi-modal\footnote{If the output space is multidimensional, then this means the component-wise distributions.}. We therefore say that the available data does a poor job constraining our model at $\mathbf{x}$. 

It may seem like the denominator of Eq.~\ref{eq:parametric-bayes} is the bottleneck for large models. As the dimensionality of $\boldsymbol{\theta} \in \mathbb{R}^{N_p}$ grows, it quickly becomes impossible to compute the $N_p$-dimensional integral. While this certainly matters in some contexts, most practical algorithms can avoid explicitly computing the evidence. The real challenge is that, regardless of whether you know the normalized posterior or just the unnormalized numerator, there's no obvious way to actually sample the distribution. Put another way, there's not an easy way to guess what parameters $\boldsymbol \theta$ might be more likely to maximize $p(D|\boldsymbol{\theta})p(\boldsymbol{\theta})$ (obviously). The geometry of the $N_p$-dimensional posterior can be extremely peculiar due to variable correlations, and the \textit{curse of dimensionality}\footnote{As $N_p$ grows, the number of samples you would need to effectively explore the space grows exponentially. Thus, the ratio between the useful volume of the target distribution and the volume covered by a random sampling strategy goes to zero exponentially fast with respect to $N_p$.} ensures that any uninformed strategy for sampling parameters is going to be increasingly ineffective. Navigating this restriction is an active area of research. Some of the most prevalent approaches include Maximum-A-Posteriori (MAP) inference \citep{Lippmann1991}, Variational Inference, or Markov Chain Monte Carlo (MCMC)\citep{Jospin2022} which are all implemented as training strategies in AutoBNN. We now explain each of these methods. 

\subsection{Strategies for Sampling High-Dimensional Posteriors}
\subsubsection{Maximum-A-Posteriori Estimation}
We have actually explored this method already in Section~\ref{subsect:example-maximizing-posterior}. Recall that we called $\boldsymbol{\hat \theta}$ a Maximum Likelihood Estimator (MLE) because it maximized the likelihood term in Bayes' Theorem. However, $\boldsymbol{\hat \theta}$ was also the \textit{Maximum-A-Posteriori} (MAP) Estimator--the parameter that maximized the posterior. In general, the prior is not a constant function so the MLE is not generally the same as the MAP Estimator. Take a look at Fig.~\ref{fig:posterior-visualizatoin}(b). Recall the interpretation of the negative log posterior as a cost function. One could repeatedly initialize different MAP fits and identify the different local minima they land in. Doing repeated MAP estimates of a neural network's parameters is essentially the same as training a neural network many times with different initializations and using their collective predictions to estimate model uncertainty. MAP estimators, shown by black dots, can represent different modes of the posterior but reflect almost no structure about the posterior beyond this. In general, the histogram of modes will not fully reflect the local volume of each mode either meaning that easy-to-find but unimportant modes may be sampled more frequently than they should. This is one of the least computationally taxing strategies to explore the posterior, but it yields one of the least faithful representations of the posterior.

\subsubsection{Variational Inference}
Variational methods involve guessing a surrogate solution $q_\phi$ parameterized by some parameters $\phi$ and varying $\phi$ until the solution $q_\phi$ meets some criteria indicating they are a good approximation to the actual solution you're after. In many-body physics, $q_\phi(\mathbf{x})$ can be a guess of the form of the many-body wavefunction that is the eigenfunction of the Hamiltonian $\hat H$ with the lowest energy $E_0$. One varies $\phi$ until the quantity
\[
E[q_\phi]
=
\frac{\langle q_\phi|\hat H|q_\phi\rangle}
{\langle q_\phi|q_\phi\rangle},
\]
is minimized, according to a theorem claiming that any trial function will have $E[q_{\boldsymbol \phi}] \geq E_0$.

Variational inference guesses a parameterized distribution $q_{\boldsymbol \phi}$ with the goal of varying $\boldsymbol \phi$ until $q_{\boldsymbol \phi}$ is as close as possible to $p(\boldsymbol{\theta} | D)$. In particular, this surrogate distribution $q_{\boldsymbol \phi}$ should be significantly easier to sample parameters $\boldsymbol \theta$ from. Ideally, the dimension of $\boldsymbol{\phi}$ is small, but $q_{\boldsymbol \phi}(\boldsymbol \theta)$ is still expressive enough to approximate $p(\boldsymbol{\theta} | D)$. Fig~\ref{fig:posterior-visualizatoin}(c) sketches out the concept of optimizing to find the optimal parameterization $\boldsymbol{\phi_*}$. Similarity between distributions is often measured with the Kullback-Leibler divergence (KL) as 

\begin{equation}
\begin{aligned}
\mathrm{KL}\!\left(
q_{\boldsymbol{\phi}}(\boldsymbol{\theta})
\,\Vert\,
p(\boldsymbol{\theta}\mid D)
\right)
= \mathbb{E}_{q_{\boldsymbol{\phi}}(\boldsymbol{\theta})}[\log q_{\boldsymbol{\phi}}(\boldsymbol{\theta})] - 
\mathbb{E}_{q_{\boldsymbol{\phi}}(\boldsymbol{\theta})}[\log p(\boldsymbol{\theta}\mid D)]\\
= \mathbb{E}_{q_{\boldsymbol{\phi}}(\boldsymbol{\theta})}[\log q_{\boldsymbol{\phi}}(\boldsymbol{\theta})] -\mathbb{E}_{q_{\boldsymbol{\phi}}(\boldsymbol{\theta})}[\log p(D|\boldsymbol{\theta})] - \mathbb{E}_{q_{\boldsymbol{\phi}}(\boldsymbol{\theta})}[\log p(\boldsymbol{\theta})] + C
\end{aligned}
\end{equation}

Therefore, maximizing similarity corresponds to maximizing the above up to an irrelevant constant. Often this is cast as maximizing the evidence lower bound, or ELBO, given as
\begin{equation}
    \mathcal{L}_{\mathrm{ELBO}}(\phi)
=
\mathbb{E}_{q_\phi(\theta)}
\left[
\log p(D|\theta) + \log p(\theta) - \log q_\phi(\theta)
\right]
\label{eq:ELBO}
\end{equation}

Equivalently, one may minimize the negative ELBO,
\(-\mathcal{L}_{\mathrm{ELBO}}(\phi)\).

The expectations are taken in practice using Monte Carlo integration of the $N_p$-dimensional integral. We must repeatedly perform this integral every time we want to vary $\boldsymbol \phi$. Although this seems intractable, the parameters $\boldsymbol \phi$ can be efficiently optimized by leveraging gradient based methods from deep learning to evaluate the ELBO over small batches of data. The actual variational ansatz $q_{\boldsymbol{\phi}}(\boldsymbol{\theta})$ is generally built from distributions in the exponential family. The most common choice is the multivariate normal distribution given the ease of sampling from it. We can perform some simple estimates. Suppose we use \textit{mean-field} Variational Inference, where each parameter is considered independent
\begin{equation}
    q_{\boldsymbol{\phi}}(\boldsymbol{\theta}) = \prod_{i=1}^{N_p} \mathcal N(\mu_i, \sigma^2_i)
\end{equation}
In this case, we must optimize 2 parameters $\mu_i, \sigma_i$ for every one parameter $\theta_i$ in the original network. This is interesting because we have stayed at the same order of computational complexity as if we had trained the parameters $\boldsymbol \theta$ deterministically, yet we gained a much richer object to sample many networks from. Of course this will be slowed down some by taking the expectation value in Eq.~\ref{eq:ELBO}. Moreover, the assumption of independence can be overly restrictive and place steep limits on the expressivity of the variational ansatz\footnote{Very similar to how a Slater Determinant is limited to underrepresenting particle correlations}. However, if we allow covariance between all parameters, suddenly the number of parameters to optimize grows as the square of $N_p$. So a relatively small network with $N_p=$ 1 million parameters will now get mapped into an $\mathcal{O}(10^9)$ parameter optimization problem which no longer feels like a worthwhile trade. A spectrum of intermediate options does exist via low-rank covariance approximations or block-diagonal covariances.


\subsubsection{Markov Chain Monte Carlo}
There are many Monte Carlo methods for Bayesian estimation, but not all of them converge quickly enough when trained to BNNs to be practical e.g., Gibbs Sampling \citep{Hoffman2014}. The original Metropolis algorithm \citep{Metropolis1953} is particularly useful, however,  as it avoids computing the marginal distribution $p(D)$ that makes computing the posterior restrictive. Instead, it relies on constructing a chain of samples that eventually reach an equilibrium distribution by accepting or rejecting adjacent samples based on their relative probability. This ratio is independent of the marginal, but samples can often take a while to converge to the target distribution \citep{Hoffman2014}. 

Considerable acceleration was obtained with the development of Hamiltonian Monte Carlo which recast the problem of drawing samples from a distribution to continuously exploring parameter space by integrating effective Hamiltonian equations of motion \citep{Duane1987}. This considerably sped up the sampling process by avoiding lengthy explorations of parameter space; however it required access to gradients of the posterior distribution which are not always available. 

However, for neural networks, the gradients of the log posterior with respect to its parameters can indeed be computed via automatic differentiation, the elegant algorithm that enables the training of most AI algorithms today by exactly computing the dependencies between parameters and tracing the computational graph through layers of the model. Taking advantage of this, the No-U-Turn Sampler (NUTS) was developed to avoid the laborious hyperparameter tuning of the HMC algorithm and to avoid retracing old paths through parameter space \citep{Hoffman2014}. Today, leading statistical ML libraries contain an implementation of the NUTS algorithm for efficiently training BNNs and this is the underlying algorithm used by AutoBNN for MCMC inference.

\begin{figure}[htbp]
    \centering
    \includegraphics[width=0.9\linewidth]{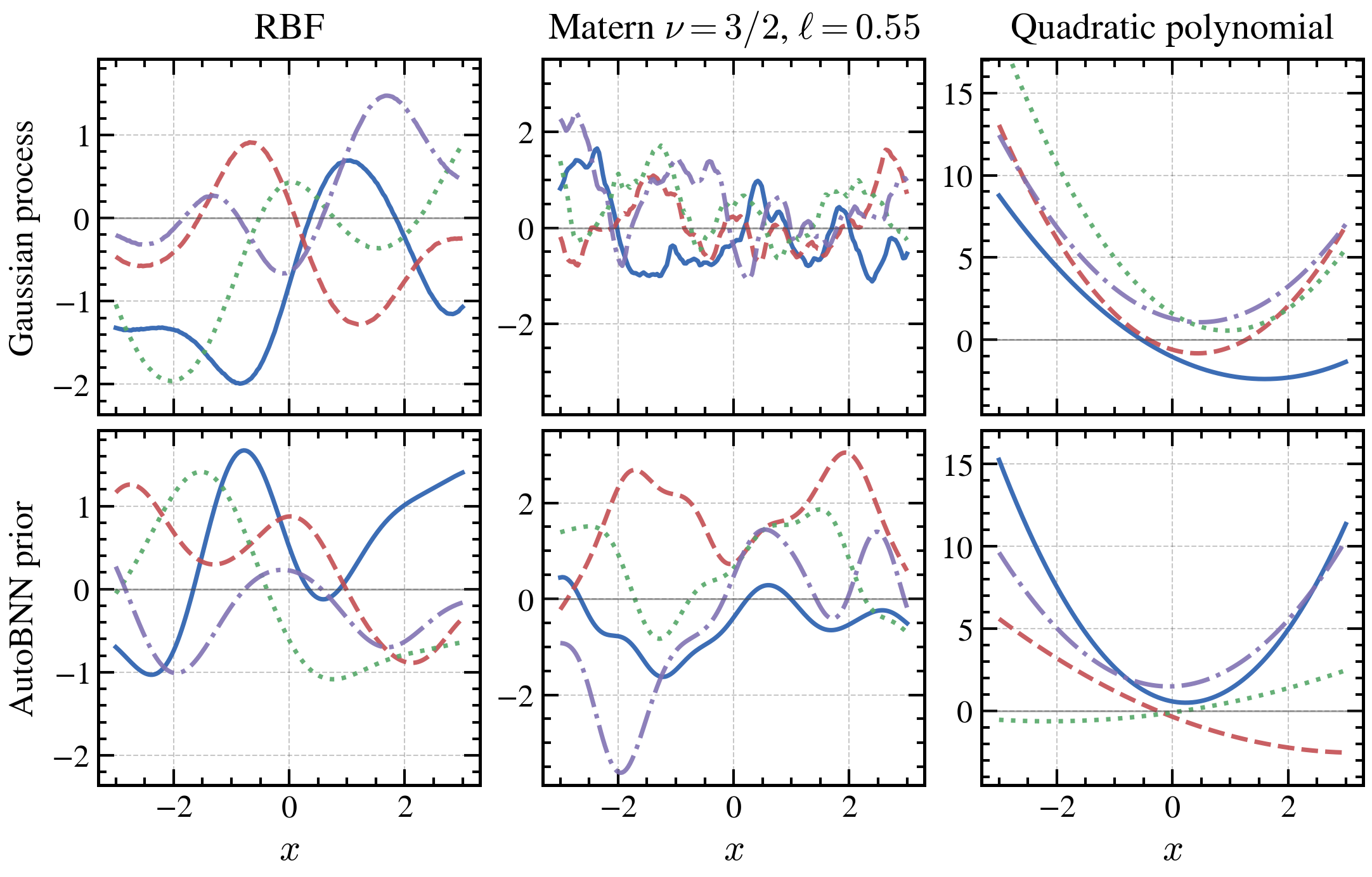}
    \caption{Representative samples from Gaussian process models with common kernels and their corresponding Bayesian Neural Network surrogates.}
    \label{fig:bnn-priors}
\end{figure}
\section{AutoBNN}

AutoBNN \citep{AutoBNN2024} is a software tool released by Google for the purpose of time series forecasting. It builds upon the close ties between Gaussian processes and Bayesian Neural Networks. 

Suppose we built a neural network $f_{\boldsymbol \theta}(\mathbf{x})$ and tried to randomly initialize its weights several times before training them via $\boldsymbol{\theta}_i \sim p(\boldsymbol \theta)$. This is the idea of a Bayesian Neural Network, and there are interesting questions we could ask before we ever train them to data. For example, what would the set of functions $\{f_{\boldsymbol \theta_i}\}$ look like initially? What if we chose different activation functions or sampled the weights from different distributions? 

In fact, it has been known for 30 years that a very wide neural network initialized with random weights from any distribution will approach a Gaussian process \citep{Neal1996} in the limit of infinite width.
More specifically, the $\{f_{\boldsymbol \theta_i}\}$ will look very similar to the functions that could have been sampled from a Gaussian process with some well-chosen mean and kernel functions. Those mean and kernel functions are not explicitly given, but this result just says those functions do exist and are dependent on the choice of activation functions and $p(\boldsymbol \theta)$. Therefore, the choice of activation functions and weight initialization strategy collectively induce a \textit{functional prior} on neural networks.

More recent research has explicitly determined which activations and parameter priors will yield networks with the same functional priors as GPs with well-known kernels \citep{Meronen2021}. Fig.~\ref{fig:bnn-priors} illustrates this correspondence. The top row demonstrates what types of functions can be drawn from different GPs with common kernels. The bottom row demonstrates what sampled neural networks with these activations and weight distributions might look like. Just like GPs, these networks can be combined with analogues of kernel addition, multiplication, and concatenation to support functional priors identical to composite GP kernels \citep{Pearce2019}. 

A great advantage of Gaussian processes is the flexibility of choosing what sort of kernel, or covariance function, best describes the data. Combining kernels via addition or multiplication can lead to even more expressive GPs, but without sufficient knowledge of the data or an existing idea of its structure, finding the correct kernel can be challenging. Some work has been done to automate the process of kernel construction \citep{Duvenaud2014} in Gaussian processes, but Gaussian processes are still limited by their poor scaling to training data. The main computational restriction of GPs is the required inversion of a matrix, an $O(N^3)$ operation. In contrast, training neural networks is closer to linear scaling with the number of data points. Additionally, the closed form expression for a conditioned GP is dependent on the assumption of a normal likelihood of observations. This assumption can be relaxed but it is challenging to find popular implementations that efficiently function out-of-the-box.

AutoBNN combines these many ideas with the intention of inheriting the flexibility and expressivity of Gaussian processes via BNN analogues, while taking advantage of the much more favorable computational scaling with respect to large datasets and the considerable work done to accelerate deep learning inference and training via GPUs. It provides multiple pre-built architectures that essentially enable a model to discover which compositional kernel analogues best describe the available data. 

We utilize the AutoBNN framework to explore more sophisticated statistical models of fusion cross section data. The rest of this chapter discusses characteristics of pre-designed models and how they perform on some of the same tests carried out in Chapter 2. To paraphrase \citep{Blei2017}, MCMC is better suited than Variational Inference to situations where we have limited data and have the computational resources to obtain very precise samples. In this work, our datasets are sufficiently small and the computational resources we use are considerable enough to train the AutoBNNs via MCMC.

\section{Methods}
 \begin{figure}[htbp]
   \centering
\includegraphics[width=0.9\textwidth]{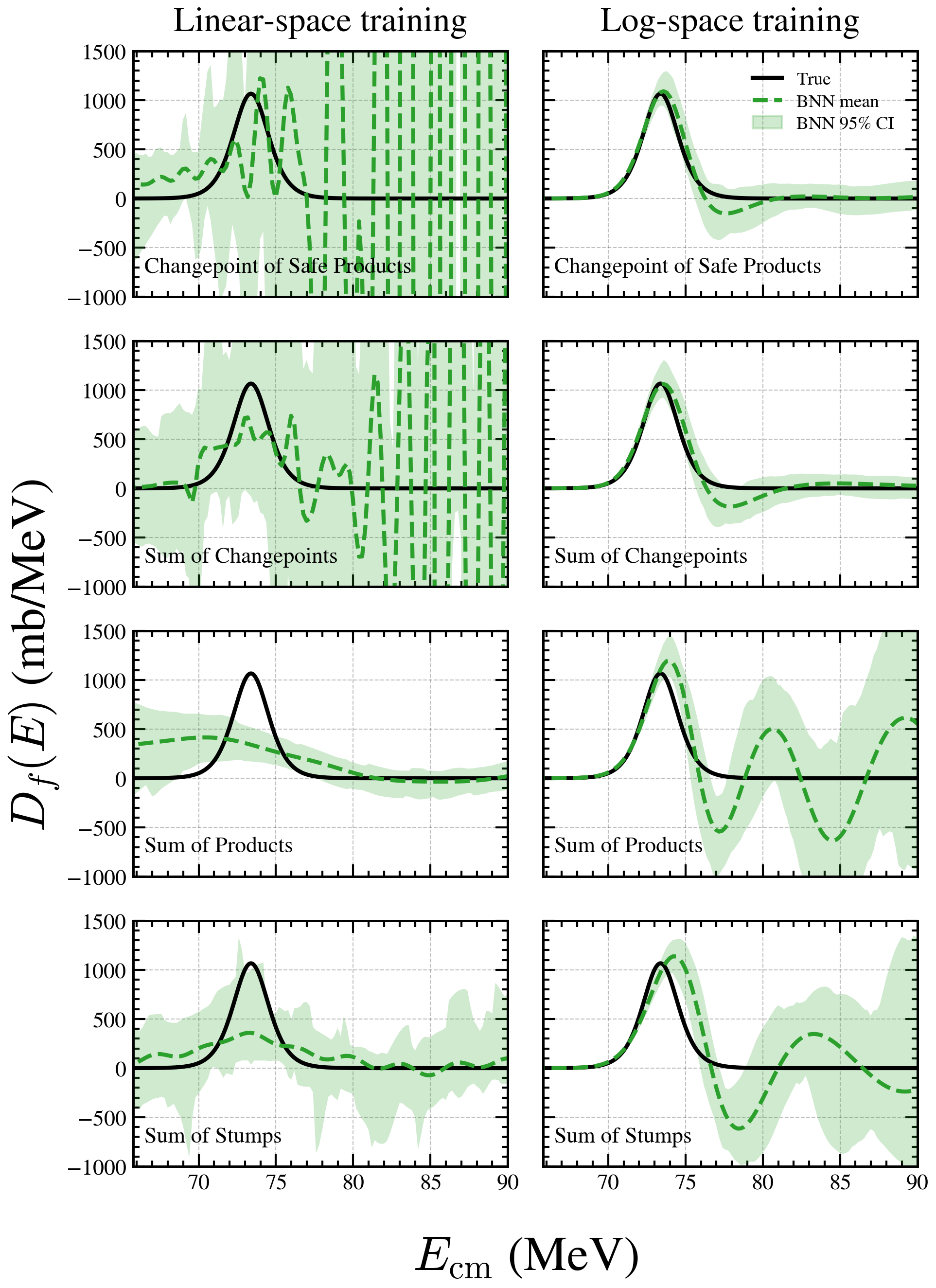}
  \caption{Left column: Training AutoBNN models in linear space. Right column: Training in log space. All data was generated at a noise ratio $\eta=10^{-2}$ with sparse data at $\Delta E=1.75$ MeV.}
   \label{fig:log_vs_autobnn}
    \end{figure} 

To accommodate the computational challenges of training AutoBNN models via MCMC, we utilized High Performance Computing Clusters (HPC) at the Institute for Cyber-Enabled Research at Michigan State University. Considerable time was spent modifying the source code of the AutoBNN package to deploy it across GPUs, as the package is no longer regularly maintained. The modified version is publicly available at \url{https://github.com/aaron-philip/autobnn-barrier/tree/main}. We developed shell scripts that navigated the HPC Linux file system, could read in custom configurations for model architecture, datasets, and parallelization strategies and automate the submission of compute jobs to the SLURM job submission manager. This required writing modular code flexible enough to handle testing of several ideas. This code is publicly available at \url{https://github.com/aaron-philip/barrier-extraction/tree/main} and relies on the aforementioned version of AutoBNN. It can be deployed with minimal effort on other computing systems. 

AutoBNN developers have designed certain architectures that explore several possible composite BNNs mirroring compositional kernels in GPs. This includes the ``Sum of Products, Sum of Stumps, and Sum of Stumps and Products'' architectures that explore weighted sums of simple combinations of kernels as well as the ``Changepoint of Products'' or ``Sum of Changepoints'' that can combine different kernels based on the data region one is in rather than fixing the kernel on the whole domain. 

All models were trained via the NUTS algorithm with 4 chains and 1000 burn-in steps. Using 4 chains allowed the model to explore a few different modes of the posterior and reduce the chance of obtaining samples that get ``stuck'' and are unable to reach relevant modes of the posterior. We empirically found 1000 burn-in steps to be sufficient to converge to a stationary distribution that mirrors the posterior. Because these models are neural networks, there is no concern of numerical noise entering the derivative computation as automatic differentiation allows us to compute exact derivatives. 

\section{Results}
Early testing on very high quality data from the Wong toy problem showed that only the above-mentioned custom architectures were robust in reproducing the data. In addition, there was a considerable difference in the quality and utility of models trained to data in linear space vs logarithmic space. Figure \ref{fig:log_vs_autobnn}
demonstrates the superior performance of models trained in logarithmic space in the right column as compared to their counterparts trained in linear space. Even though the data is much sparser than typically encountered, the Changepoint models trained in logarithmic space capture the barrier everywhere far more effectively although they likely underestimate their uncertainties at high energies. In fact, they seem to avoid the aliasing that the Gaussian processes suffered although this is still present in many of the other models. Nevertheless, training in log space then transforming back to linear space come inference time proves to be an effective regularization step. All BNNs are trained in logarithmic space for the remainder of this thesis. Although a few different prior likelihoods were tested, the normal likelihood with a prior on the noise scale arising from a lognormal distribution was the most effective likelihood model.

 \begin{figure}[htbp]
   \centering
\includegraphics[width=0.9\textwidth]{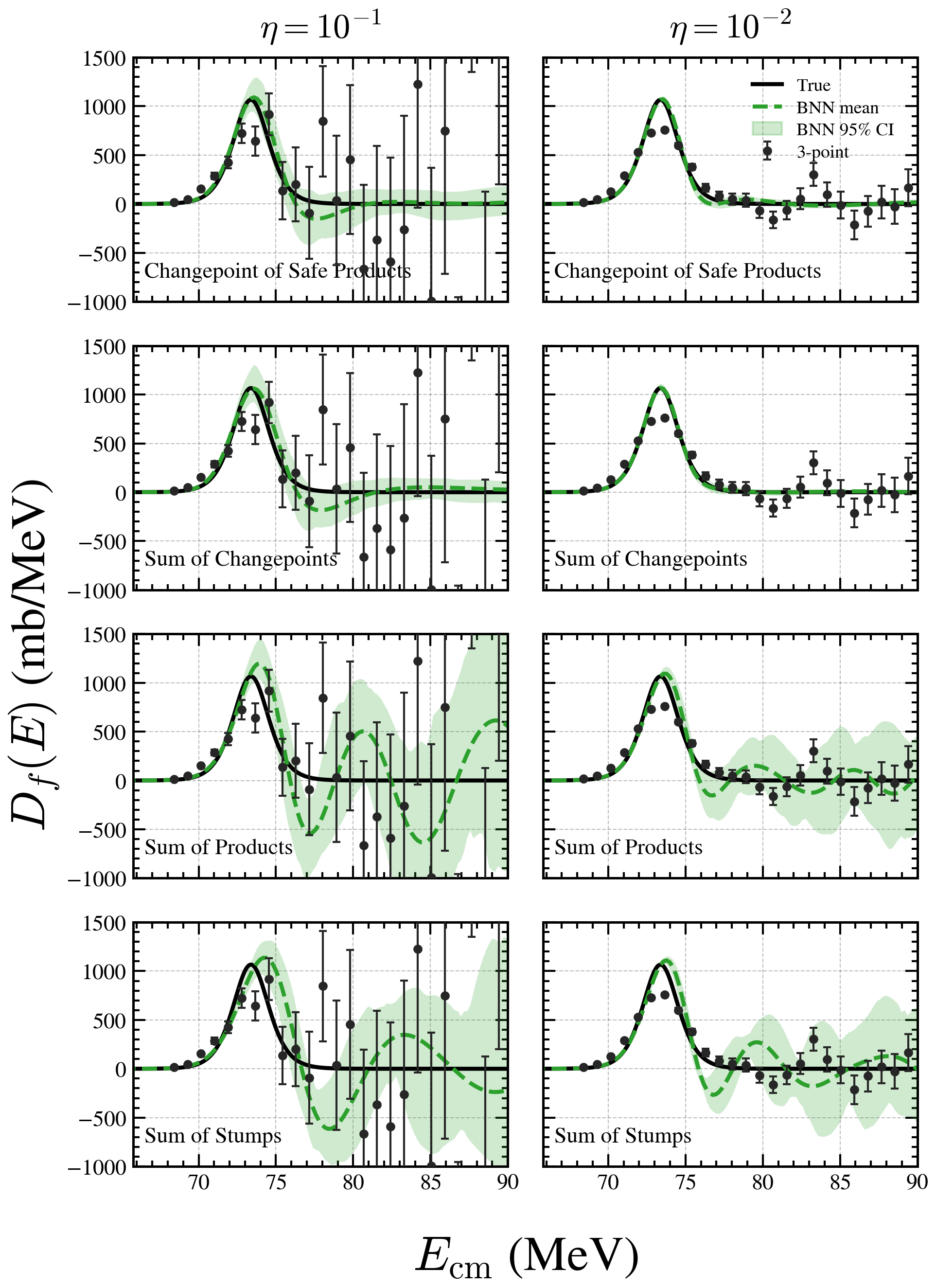}
   \caption{Left column: Noise ratio $\eta=10^{-1}$. Right column: Noise ratio $\eta=10^{-2}$ for a variety of AutoBNN architectures trained to $\log\sigma(E)$. $\Delta E = 0.88$ MeV for all models. 3-point spacing uses $\langle \Delta E \rangle= 2.63$ MeV.}
   \label{fig:noise_ratio_autobnn}
 \end{figure}

 \begin{figure}[htbp]
   \centering
\includegraphics[width=0.9\textwidth]{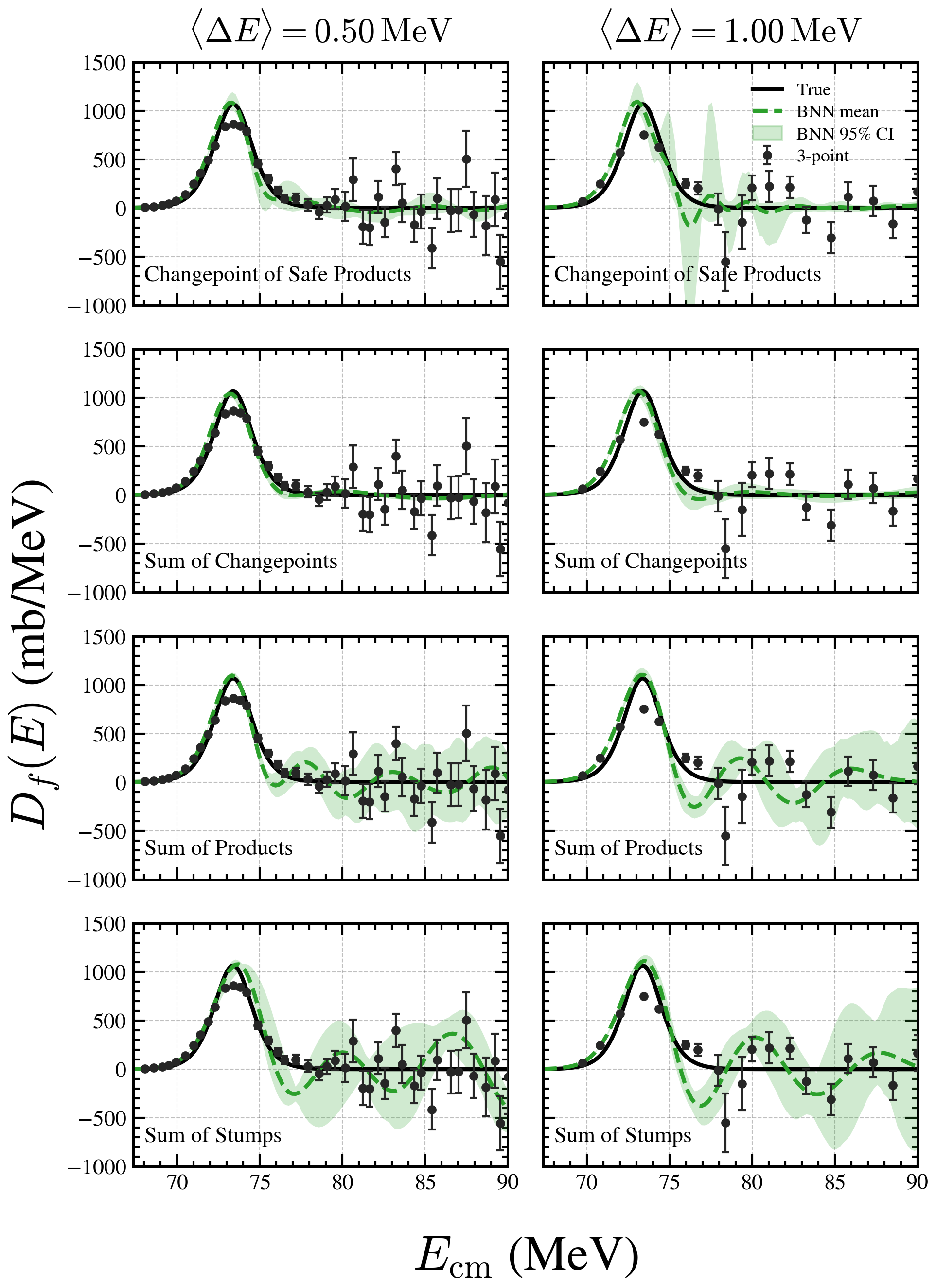}
   \caption{Left column: Spacing $\Delta E=0.5~\mathrm{MeV}$. Right column: Spacing $\Delta E=1.0~\mathrm{MeV}$. Both noise levels were fixed at  $\eta=10^{-2}$ and models were trained to $\log\sigma(E)$. Data irregularity is defined by a coefficient of variation $\kappa=0.50$. 3-point spacing uses $\langle \Delta E \rangle= 2.03$ MeV.}
   \label{fig:data_irregularity_autobnn}
 \end{figure}
 
\subsection{Effect of Noise}
Figure \ref{fig:noise_ratio_autobnn} demonstrates how a variety of AutoBNN models perform under the same two noise level choices tested in Chapter 2. The Changepoint models perform very impressively at high energies, capturing the true distribution very well and not blowing up their predictions and uncertainties. The other two models, like the Gaussian processes, seem to be sensitive to the type of noise they are trained to: they capture the barrier very well and make reasonable predictions of uncertainties at higher energies, but this collapses into aliasing when the noise level is higher. 

\subsection{Data Irregularity}

Interestingly, one of the Changepoint models in Figure \ref{fig:data_irregularity_autobnn} exhibits evidence of an unconverged chain. Despite the irregular data, all the models except the Sum of Stumps model capture the barrier characteristics well just like the Gaussian process did. The Sum of Stumps model seems to contain several distinct modes in the case of having trained to the data with mean step size $\Delta E=0.5$ MeV. It would be interesting to follow up with a study of whether this is a desirable effect, or one that goes away with more chains or burn-in samples per chain. 

\section{Conclusions}
The AutoBNN architectures are capable of exploring some varied and expressive models. Some of the main takeaways are that \begin{enumerate}
    \item Training in logarithmic space is much more effective for training to cross section data. 
    \item The Changepoint models seem capable of handling high energy components of the distribution function very well when their MCMC chains are converged. 
    \item The remaining models often have the same behaviors as Gaussian processes such as the periodic aliasing that occurs at higher energies and the sensitivity to noise levels. 
\end{enumerate}

It may be worth exploring what each AutoBNN model looks like when translated back into a compositional kernel. Several of the AutoBNN models give nearly identical results to the GPs in Ch. 2. Overall, Gaussian processes perform similarly well for a fraction of the computational cost used to train those BNNs. This is likely due to the two advantages a GP has over a BNN: the ability to specify an arbitrary mean function and then simply condition on data to learn ``corrections'' and the ability to explicitly encode observational uncertainties when training the GP. Nevertheless, the Changepoint models are significantly more robust than both GPs and other BNN models in the high energy regime and avoid the spurious aliasing every method thus far has been susceptible to. A future experiment might test the limits of the secondary structure that Changepoint models can extract from data with a more complicated toy model than the single barrier Wong formula used here.  

\clearpage
\chapter{Extracting Barrier Distributions from Fusion Cross Sections}\label{c:chap4}
\chaptermark{}

\section*{Abstract}

In this chapter, we train AutoBNN models to the same datasets used in \cite{Godbey2024} and report the extracted barrier distributions. We then analyze experiments for which no quoted measurement uncertainties were available. We provide novel insights into the nature of the true barrier distributions by 

\begin{enumerate}
\item Considering which features of the barrier our ensemble of conditioned models agree on.
\item Taking into account the analysis from the previous two chapters to determine when a Gaussian process model or the 3-point formula may be unlikely to accurately represent the distribution.
\item Leveraging the continuous posterior available to us to interpolate features of the barrier.
\end{enumerate}

Although we already know that the 3-point discrete difference formula can be unreliable, we see that its discrete nature is often insufficient to fully understand the barrier. As one expects, high quality experimental data enables our models to confidently represent important characteristics of the barrier distribution. Considering an ensemble of models also enables us to make experimental recommendations. The energy ranges in which several models disagree or quote high uncertainty specify where more experimental cross section data is needed. 

\section{Results}
We display predictions for the barrier distribution made with the 3-point formula, a Gaussian process model with a Radial Basis Function kernel, and the two Changepoint AutoBNN models that excelled on the representative benchmark problem in Ch. 3. 
\begin{figure}[p]
    \centering

    \includegraphics[width=\textwidth]{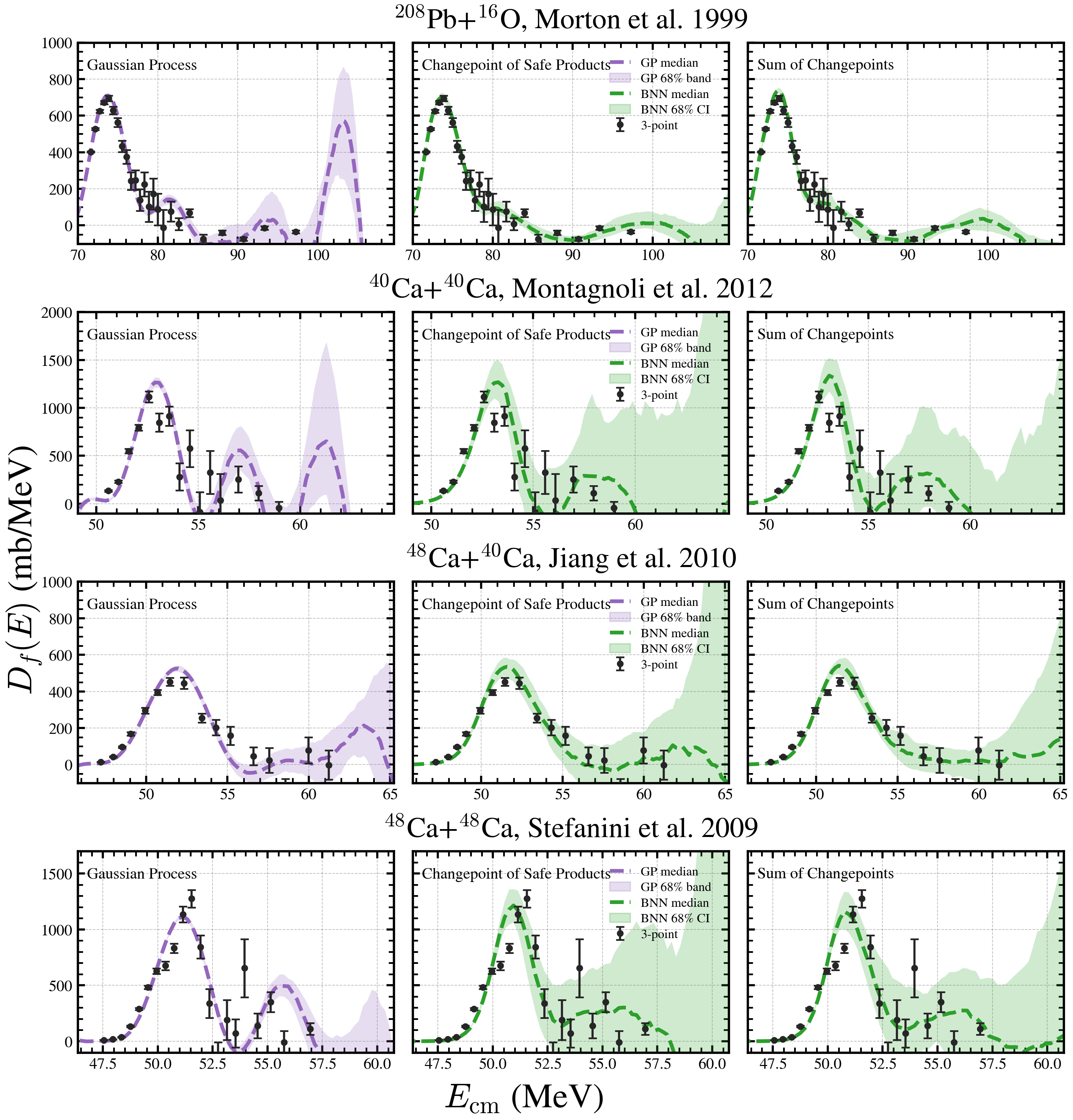}

    \caption{
    Fusion barrier distribution results for real experiments in \citep{Morton1999, Montagnoli2012, Jiang2010, Stefanini2009}.
    Each of these experiments provided experimental estimates of their uncertainties that the Gaussian process and 3-point formulas directly leveraged for their predictions. 
    }
    \label{fig:barrier-page1}
\end{figure}
\subsection{Experiments with Quoted Measurement Uncertainties}

\subsubsection{$^{16}$O+$^{208}$Pb}
The data shown in Fig.~\ref{fig:barrier-page1}(a) was the highest quality dataset of all fusion cross-section results considered. Thus, it makes sense that the models considered are mostly consistent up to 100 MeV. All models collectively identify a secondary structure around 80 MeV that plateaus before falling to zero. Moreover, the two continuous models both indicate a tertiary rise in the barrier around 100 MeV that would be impossible to notice with just the discrete 3-point model. Given that the Changepoint models generally performed best at high energies, it is likely that the Gaussian process is largely aliasing the height of the peak near 100 MeV, however.

\subsubsection{$^{40}$Ca+$^{40}$Ca}
While all models agree in Figure \ref{fig:barrier-page1}(b) up to the peak of the barrier, the only conclusion that considering the committee of models seems to give is that the barrier is poorly constrained after 56 MeV. Both continuous models predict a rise in the barrier around 58 MeV but the wide uncertainty bands indicate a high degree of uncertainty. Note that the secondary rise would be impossible to even see using the discrete 3-point formula. Moreover, as shown in Ch. 2, the GP often exhibits overconfident predictions which may explain the discrepancy between the size of the BNN uncertainties and the GP uncertainties. Therefore, further measurements in the energy range of [57,60] MeV will be vital to pinning down the structure of the barrier. Again, note that this very specific recommendation would be much more challenging to give based solely on either the GP or 3-point formula's predictions. 

\subsubsection{$^{48}$Ca+$^{40}$Ca}
In Figure \ref{fig:barrier-page1}(c), the Gaussian process regression results agree very well with the Changepoint results at nearly all energies--this suggests that the experimental data is high quality enough to constrain the barrier up to around 60 MeV. Interestingly, both the Bayesian inference methods predict that the primary barrier height is larger than the 3-point formula suggests. Likewise, the AutoBNN and 3-point models are consistent with their predictions of the decline of the barrier at 55 MeV while the GP is inconsistent with the 3-point prediction.  

\subsubsection{$^{48}$Ca+$^{48}$Ca}
Of all datasets considered, we found that the experiment shown in Fig.~\ref{fig:barrier-page1}(d) was most challenging to analyze. In general, the primary peak's shape, height, and location are typically quite reliable when extracted with the 3-point formula. In this case, none of the shape, height, or location of the primary peak agree with the two Bayesian inference methods. We find that the measurement around 53 MeV may be inconsistent with the rest of the measurements. Furthermore, the GP indicates a confident prediction that a second peak exists around 56 MeV. However, the AutoBNN models quote a high degree of uncertainty in this region and suggest that a plateau at most exists from 52 MeV to 57 MeV. To settle this question, more experimental measurements around 55 MeV would be maximally impactful.

Again, given the lack of model consistency, it is hard to draw any conclusions about what the true barrier distribution looks like other than the fact that a secondary peak around 56 MeV is indeed plausible. However, the uncertainties of the BNNs suggest that more experimental data after 53 MeV is needed to resolve the question conclusively.
\begin{figure}[htbp]
    \centering

    \includegraphics[width=0.9\textwidth]{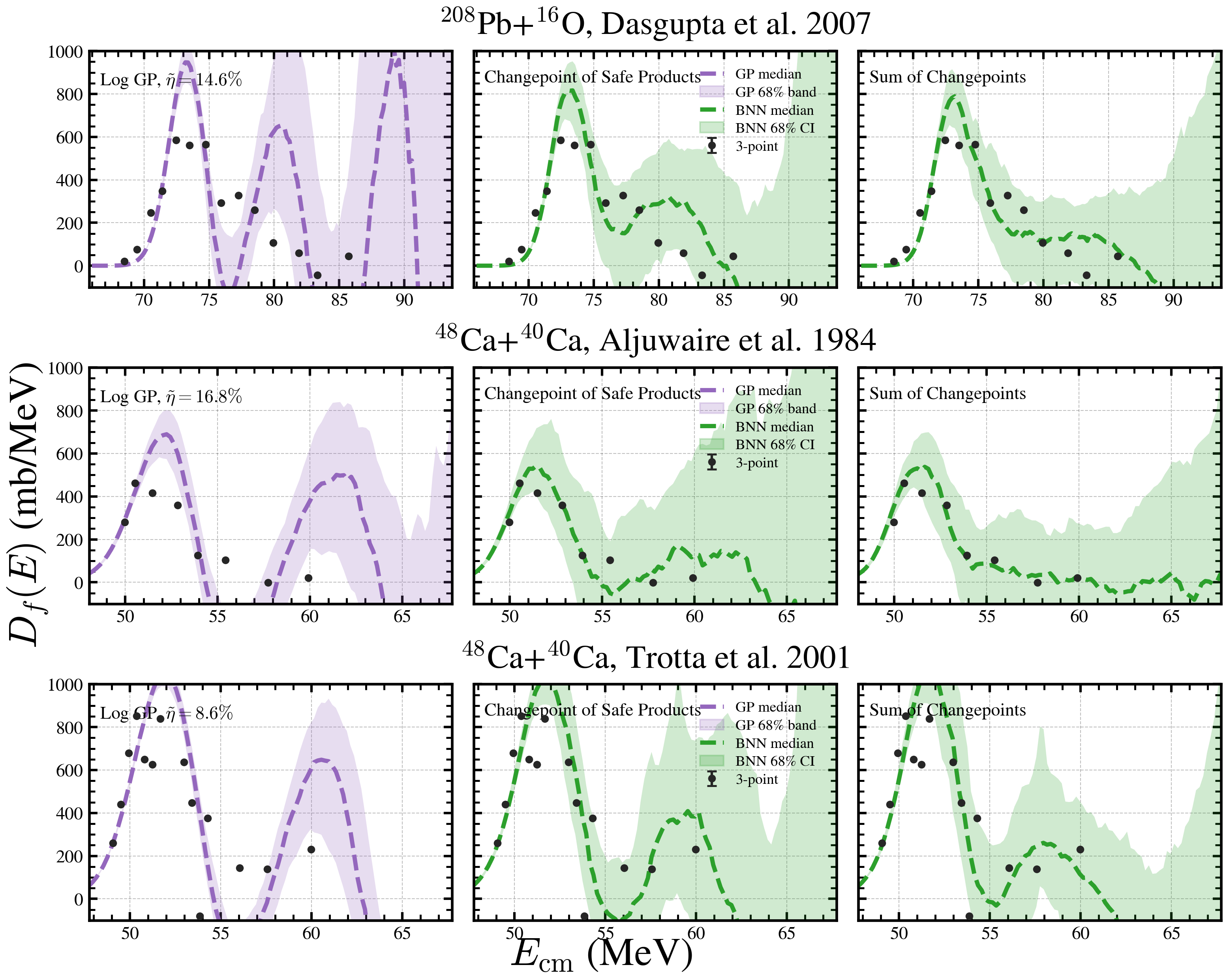}

    \caption{
    Fusion barrier distribution results for fusion experiments in \citep{Dasgputa2007, Aljuwair1984, Trotta2001}.
    Experimental uncertainty estimates were unavailable for these experiments, therefore we utilized the method introduced in Ch. 3 to infer a learned relative uncertainty $\tilde{\eta}$.
    }
    \label{fig:barrier-page2}
\end{figure}

\subsection{Experiments without Available Measurement Uncertainties}

\begin{figure}[htbp]
    \centering

    \includegraphics[width=0.8\textwidth]{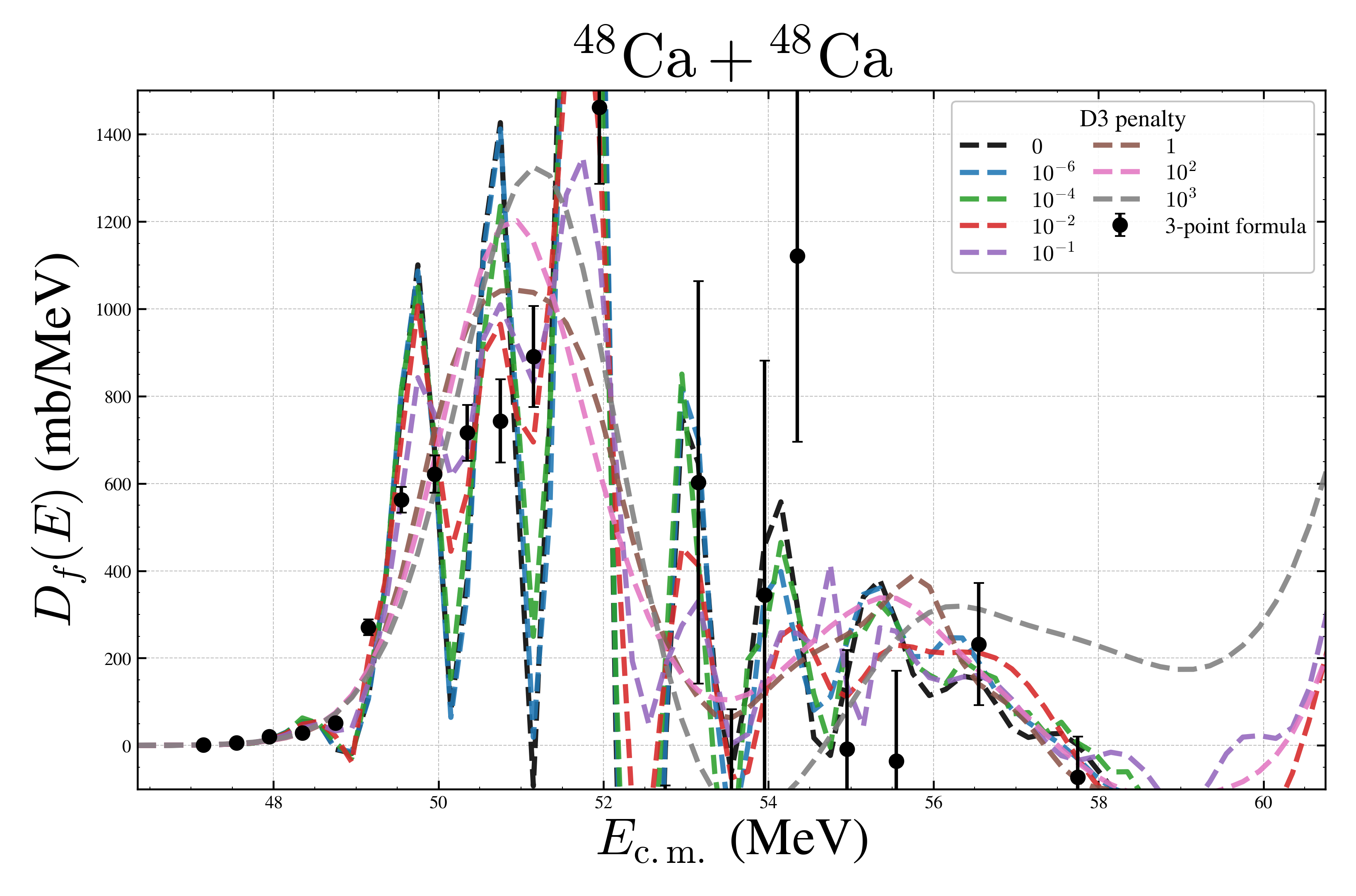}

    \caption{Repeated trains to the data in
    \citep{Stefanini2009} utilizing different relative strengths of the derivative regularizer. As the strength increased past around $\lambda=1.0~\mathrm{MeV^6}$, the barrier distribution becomes smoother.
    }
    \label{fig:stefanini-d3sweep}
\end{figure}

These datasets are quite interesting to analyze considering that we have some baseline comparisons in the previous section for each experiment. The effects of different normalizations become clear here which reflects the different techniques used to measure fusion cross-sections by experimentalists. Moreover, we cannot utilize information about measurement uncertainties even though the 3-point and GP methods both rely on it. Therefore, we employ the method of assuming a relative uncertainty and inferring it to maximize marginal likelihood as introduced in Ch. 2 for the GP and plot the central prediction of the 3-point formula without uncertainties.  

\subsubsection{$^{208}$Pb+$^{16}$O}
Fig.~\ref{fig:barrier-page2}(a) demonstrates that the GP aliases large peaks beyond the primary peak. This is expected. As shown in Ch. 2, when training is performed in logarithmic space, the primary peak is captured more reliably at the cost of enhanced aliasing above the Coulomb barrier. In this case, the Sum of Changepoint model seems most closely aligned with the results from \citep{Morton1999} in Fig.~\ref{fig:barrier-page1}(a) whereas the Changepoint of Safe Products method suggests a secondary barrier may exist. All three Bayesian methods predict a barrier that is considerably higher than the data in Fig.~\ref{fig:barrier-page1}(a) suggests. 

\subsubsection{$^{48}$Ca+$^{40}$Ca}

Both experimental datasets in this situation have substantially different normalizations which can be seen when plotting the actual excitation functions. Interestingly, nearly every method apart from the Sum of Changepoint in Fig.~\ref{fig:barrier-page2}(b) hints at the presence of a secondary peak around 60 MeV. However, the result in Fig.~\ref{fig:barrier-page1}(c) implies that no such secondary peak exists. The BNN generally quotes large uncertainties in the above-barrier region indicating low confidence in the prediction. Again, we stress the importance of this feature in the BNN as a utility that enables us to avoid claiming evidence of nuclear structure based on noise, insufficient data, or models with poorly calibrated uncertainties.

\subsection{Note: A Useful Prior}

\begin{figure}[htbp]
    \centering

    \includegraphics[width=0.9\textwidth]{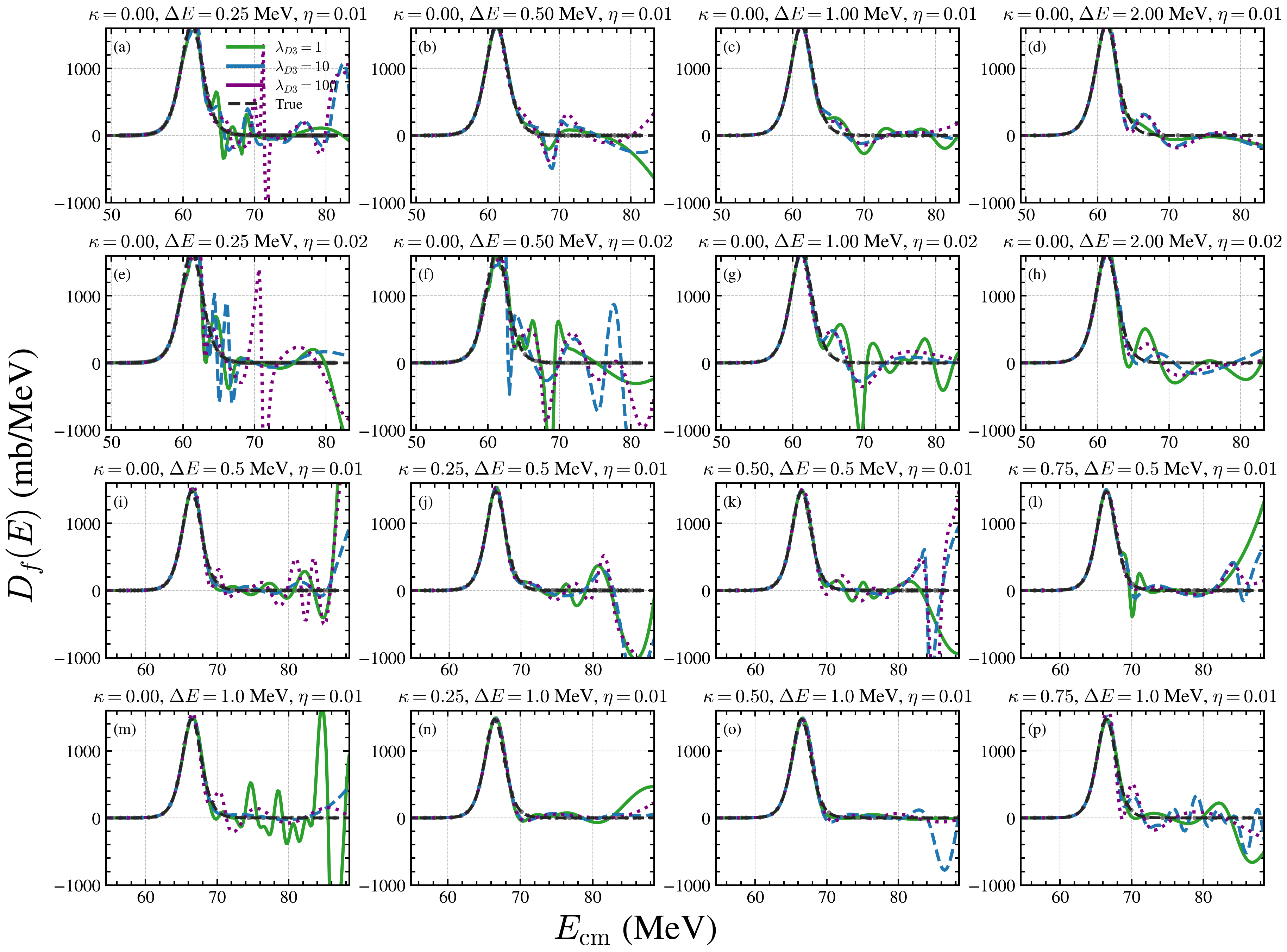}

    \caption{Sum of Changepoint mean predictions on diverse simulated Wong formula data. The strength of the penalty does not universally imply that discontinuities will vanish. }
    \label{fig:16-datasets}
\end{figure}

While training our Changepoint BNNs, we found that many times, the second derivative could become numerically unstable at the learned location of a changepoint. Occasionally, some of the less smooth BNN kernels exhibited derivatives that were particularly noisy and nearly discontinuous. To compensate for these situations, we implemented a prior/regularizer that penalized the size of the third derivative 

\begin{equation}
\begin{aligned}
    p'(\theta) \propto 
    p(\theta)
    \exp(-\lambda L[g_{\theta}]),
    \\
    L[g_{\theta}]
    &=
   \frac{1}{E_2-E_1} \int_{E_1}^{E_2}
    \left[
    \frac{d^3 g_{\theta}}{dE^3}
    \right]^2
    dE .
\end{aligned}
\end{equation}

Here, $\lambda$ controlled the relative strength of the prior. Note that this is equivalent to adding a term to a cost function as discussed in Ch. 3. Fig.~\ref{fig:stefanini-d3sweep} displays several different choices of $\lambda$ and how they impacted the extracted barrier distribution. 

To tune the strength of $\lambda$ in a more rigorous way than the eye test, we trained the Sum of Changepoint BNNs with several choices of $\lambda$ to several diverse Wong curves and identified which penalty strengths best enabled the BNN to recover the Wong curve within uncertainties on average. We found $\lambda=10.0~\mathrm{MeV^6}$ to be the best choice and all Changepoint models in this chapter were trained with this penalty strength. However, observe the Sum of Changepoint model predictions in Fig.~\ref{fig:16-datasets}. One can see that even a strong penalty does not necessarily remove discontinuities. In short, the choice of penalty strength may require nuance and expert guidance to select. Obviously, if one sees the discontinuity of the purple prediction in the top-left panel of Fig.~\ref{fig:16-datasets}, it is clear that a rerun or different choice of strength may be necessary.

\clearpage
\chapter{Conclusions}\label{c:chap5}
\chaptermark{}

In this thesis we tackled a central problem in understanding the fusion of heavy ions: extracting a barrier distribution from experimental measurements with well-quantified uncertainties. Barrier distributions rely on a numerically unstable second derivative of discrete data and are therefore hypersensitive to measurement noise and differences in any sort of interpolation scheme. Standard methods using point difference estimates are susceptible to these issues and can only offer estimates of the barrier distribution at discrete locations, making it challenging to interpret finer features of the barrier distribution. To address these challenges, we use Bayesian inference to assign probabilities to different continuous excitation functions that could reasonably describe the observed data. We generalized a method introduced in \cite{Godbey2024} and evaluated it against a representative benchmark problem. We then adapted the method to incorporate the AutoBNN framework, an advanced statistical machine learning tool introduced by Google \citep{AutoBNN2024}. 

\section{Summary of Contributions and Findings} 
In Chapter 2, we found that the Gaussian process regression-based method \citep{Godbey2024} often extracts a mean prediction for the barrier distribution with false oscillations above the Coulomb barrier. A theoretical analysis coupled with numerical experiments revealed the source: the covariance matrix becomes ill-conditioned due to the experimental errors spanning many orders of magnitude. The mean prediction can be shown to decompose into a sum of functions with large amplitudes that collectively cancel. The residuals of these cancellations consequently appear oscillatory. Although mapping all quantities into logarithmic space improves this problem marginally and corresponds to a more realistic error model, the Gaussian process still contains aliasing of false peaks at higher energies as evidenced using simulated cross section data. Moreover, the quoted uncertainties are underestimated, potentially leading to mistaken conclusions about the nucleus' structure. Beyond this analysis, we introduced a complementary method to apply the Gaussian process based method to situations where no experimental uncertainties are available. This method assumes that all uncertainties come from a fixed relative error and infers it by maximizing the marginal likelihood. 

In Chapter 3, we found that certain AutoBNN architectures are capable of ameliorating this issue significantly. Specifically, the ability of these \textit{Changepoint} architectures to model different energy regions with different statistical models enables them to excel at recovering the barrier distribution in both the sub-barrier and above-barrier regions simultaneously. Moreover, the estimated uncertainties are significantly more reliable than Gaussian process models and the standard point difference approximation of the barrier. We establish this by benchmarking against the same simulated data used to analyze the Gaussian process method.

In Chapter 4, we applied these findings from our controlled analysis to real experimental fusion data. We analyze fusion of several closed-shell systems and infer qualitative behaviors of the barrier distribution that are invisible when using only the discrete point difference. Given the superior performance of the AutoBNN Changepoint models, these methods take precedence when concluding what certain features of the distribution look like. Moreover, the strong coverage capability of the AutoBNN models enables us to provide specific estimates of the energy ranges where future experimental data would be useful. We even analyze experimental data for which no measurement uncertainties are readily available using our modified GP algorithm and the AutoBNN algorithm. In many cases, we found it useful to impose an additional prior encouraging function samples with continuous second derivatives. The modified AutoBNN source code is available at \url{https://github.com/aaron-philip/autobnn-barrier/tree/main}.

We have released the source code to reproduce all experiments at \url{https://github.com/aaron-philip/barrier-extraction/tree/main}. This repository further serves as a user-friendly implementation of our method for new experimental datasets. Users can upload their own experimental data into the example/ folder and easily train their own AutoBNN models. 

\section{Limitations, Future Work, and Other Discussion Points}

\begin{figure}[htbp]
    \centering

    \includegraphics[width=\textwidth]{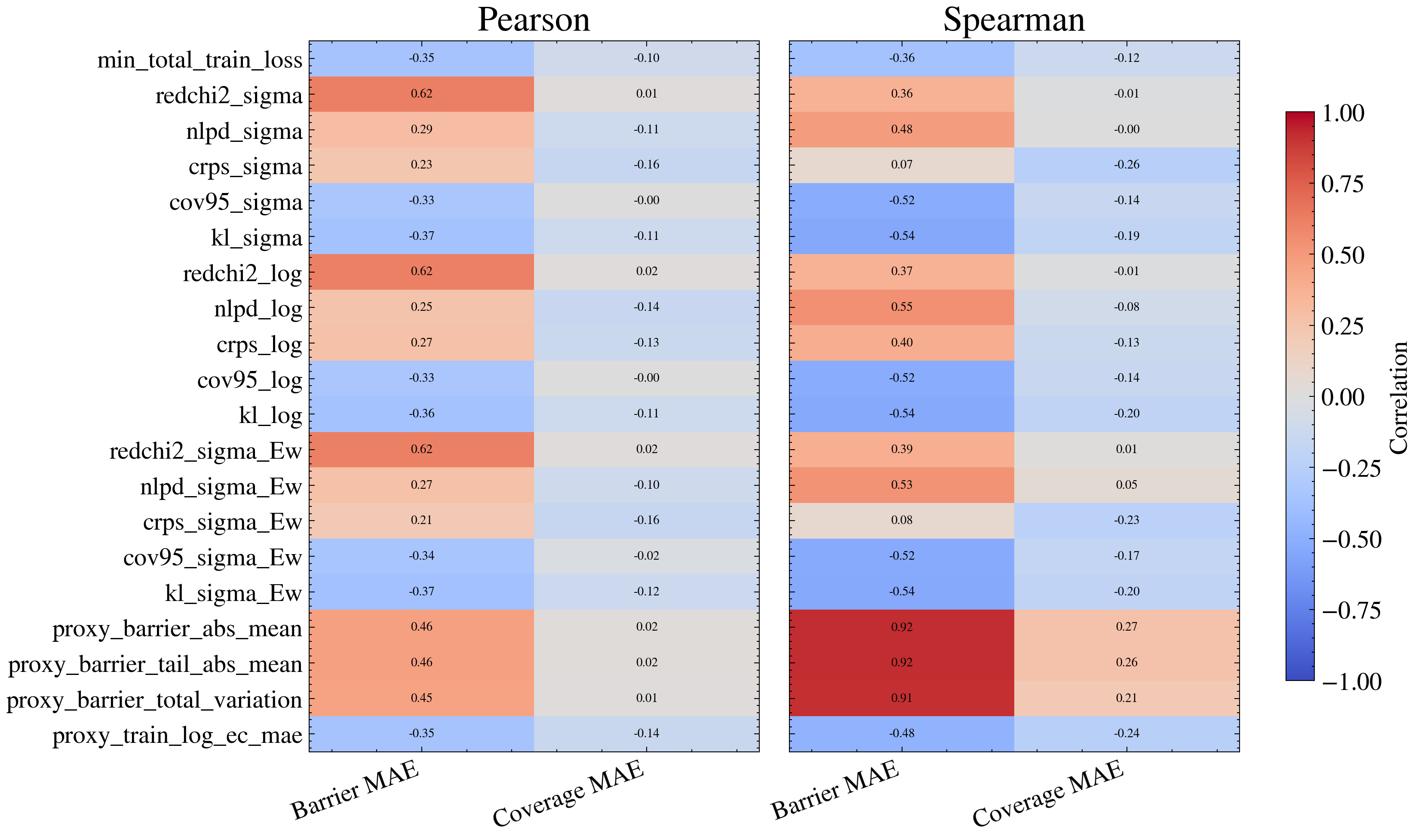}

    \caption{Many different scoring rules we experimented with and their low correlation with actual model performance. For a full description of each method, please see the linked repository.}
    \label{fig:scoring-rules}
\end{figure}

The strongest critique of the analysis in this thesis regards the Gaussian process construction. The Radial Basis Function kernel is a common kernel choice, but is very likely not the best possible kernel we could have designed and used for all numerical experiments. Indeed, the idea of Changepoint kernels was traditionally introduced in the Gaussian process context. We did indeed carry out preliminary investigations into other common kernel choices and found little improvement in results. 

However, the real purpose of using the Gaussian process with Radial Basis Function kernel was to set a baseline with which to compare the AutoBNN tool. The initial goal was to compare two \textbf{out-of-the-box} methods and see which worked better. Of course, this did not remain the case because we modified the source code of AutoBNN to implement additional experiments and the third derivative prior. Therefore, one goal of releasing the linked repository is for other users to have a tool that works out of the box. 

One way to view this thesis is in analogy to a validation set when training models. The validation set is used to calibrate \textit{hyperparameters}\footnote{parameters defining your model that are not directly learned but instead are user-adjusted} of a model. There are many degrees of freedom to the methods we use: what architecture should be used, what likelihood works, how strong should the third derivative penalty be, etc. By determining the answers to these questions on a controlled representative problem, we better understand how likely our method is going to work when deployed in an inference context. This thesis performs the work of analyzing which modeling choices work well so that users with new experimental data can spend as little time as possible tuning our method to their own data.

It must be stated that, although we conclusively answered many of the above questions, some modeling choices remain open questions. In particular, there are situations where the Sum of Changepoint and Changepoint of Products models yield genuinely different predictions or where one model is unstable but another is more reasonable. Unfortunately, neither model is conclusively more reliable than the other. Moreover, the strength of the 3rd derivative prior is contentious as shown in Ch. 4, and may need to be changed from the value of $\lambda = 10.0~\mathrm{MeV}^6$ that we used. Both of these changes are easy to implement via command line arguments but can be frustrating to experiment with given the high cost of training these models via Markov Chain Monte Carlo. Finally, it can be unclear how to pick a trained model from a collection of repeated training runs with different hyperparameters. One can use the lowest average loss\footnote{More precisely, the average value of the negative log posterior over all functions sampled.}, but this is not inherently correlated with better performance. Other scoring rules and estimators \citep{gneiting2007} were experimented with as shown in Fig.~\ref{fig:scoring-rules} but there is no scoring rule that seems strongly correlated with better-performing models. Rather than naively considering these different models on equal footing or with subjective measures, Bayesian model combination might provide a more principled approach to extract information from several different predictions \citep{Giuliani2024}. Individual models would be able to estimate their own uncertainties, and the “global model” could also indicate which models are more likely to reflect the true physics. All of the above deserve greater investigation and would be significant contributions to this method. 

Training the AutoBNN models via MCMC is expensive. Although not reported here, we investigated whether methods like Variational Inference or repeated MAP estimation could suitably approximate the performance of MCMC training, but found that MCMC training yielded significantly better results. However, the AutoBNN models are finite width approximations of the functional prior of Gaussian processes with specific kernels. Therefore, one of the most exciting future directions would be to translate the AutoBNN models \textbf{back into a GP kernel} to recover the infinite width limit of our AutoBNN model. A sketch of how this might work is to: 
\begin{enumerate}
    \item Train, e.g., a Sum of Changepoint AutoBNN model several times to diverse datasets. 
    \item Examine the weights of the weighted kernel combinations and note any regularly active kernel components or kernel combinations.
    \item Translate the few active AutoBNN kernel surrogates back into a real GP kernel. If the weights differ significantly across models but only a few kernel components are active, these can be treated like hyperparameters to optimize via marginal likelihood maximization.
\end{enumerate}

This would significantly accelerate the training procedure so long as a normal likelihood was used, and could be interpreted formally as the infinite width limit of the AutoBNN models. Of course there are many potential issues. For instance, the AutoBNN model may really need the flexibility of having all possible kernel combinations and use many different combinations for different curves. An interesting note is that the infinite width limit will kill the multi-modality of the function distribution at a fixed input location that is currently allowed because of the finite width. If this is an important factor in AutoBNN's strong performance the Gaussian process translation may not work as well as anticipated. However, one of the key advantages that GPs hold over BNNs is the ability to explicitly encode uncertainties in observed data directly into the conditioning of the GP. By contrast, BNNs must infer the underlying error structure. We conducted some preliminary experiments by constructing a BNN likelihood that explicitly considered quoted experimental uncertainties, but this strategy generally performed worse than allowing the BNN to infer the noise scale. Finally, if the likelihood is assumed to be normal, then the posterior is given in closed form for fixed hyperparameters and the only optimization is over hyperparameters. If successful, one could sample functions from the posterior after a few seconds rather than hours. We defer such an effort to a future investigation. 



\singlespace
\bibliographystyle{apj}
\bibliography{your_bib_file}

@article{Godbey2024,
        author = {Godbey, K.},
        title = {Barrier distribution extraction via Gaussian process regression},
        DOI= "10.1051/epjconf/202430601001",
        url= "https://doi.org/10.1051/epjconf/202430601001",
        journal = {EPJ Web Conf.},
        year = 2024,
        volume = 306,
        pages = "01001",
}

@article{Wong1973,
  title = {Interaction Barrier in Charged-Particle Nuclear Reactions},
  author = {Wong, C. Y.},
  journal = {Phys. Rev. Lett.},
  volume = {31},
  issue = {12},
  pages = {766--769},
  numpages = {0},
  year = {1973},
  month = {Sep},
  publisher = {American Physical Society},
  doi = {10.1103/PhysRevLett.31.766},
  url = {https://link.aps.org/doi/10.1103/PhysRevLett.31.766}
}

@article{Phillips2021,
doi = {10.1088/1361-6471/abf1df},
url = {https://dx.doi.org/10.1088/1361-6471/abf1df},
year = {2021},
month = {may},
publisher = {IOP Publishing},
volume = {48},
number = {7},
pages = {072001},
author = {Phillips, D R and Furnstahl, R J and Heinz, U and Maiti, T and Nazarewicz, W and Nunes, F M and Plumlee, M and Pratola, M T and Pratt, S and Viens, F G and Wild, S M},
title = {Get on the BAND Wagon: a Bayesian framework for quantifying model uncertainties in nuclear dynamics},
journal = {Journal of Physics G: Nuclear and Particle Physics},
}

@article{Weizacker1935,
  title = {Zur Theorie der Kernmassen},
  author = {Weizsäcker, C. F. v.},
  journal = {Z. Physik},
  volume = {96},
  issue = {7},
  pages = {431--458},
  numpages = {0},
  year = {1935},
  month = {July},
  doi = {10.1007/BF01337700},
}

@article{BetheBacher1936,
  title = {Nuclear Physics A. Stationary States of Nuclei},
  author = {Bethe, H. A. and Bacher, R. F.},
  journal = {Rev. Mod. Phys.},
  volume = {8},
  issue = {2},
  pages = {82--229},
  numpages = {0},
  year = {1936},
  month = {Apr},
  publisher = {American Physical Society},
  doi = {10.1103/RevModPhys.8.82},
  url = {https://link.aps.org/doi/10.1103/RevModPhys.8.82}
}

@article{Mayer1950a,
  title = {Nuclear Configurations in the Spin-Orbit Coupling Model. I. Empirical Evidence},
  author = {Mayer, Maria Goeppert},
  journal = {Phys. Rev.},
  volume = {78},
  issue = {1},
  pages = {16--21},
  numpages = {0},
  year = {1950},
  month = {Apr},
  publisher = {American Physical Society},
  doi = {10.1103/PhysRev.78.16},
  url = {https://link.aps.org/doi/10.1103/PhysRev.78.16}
}

@article{Mayer1950b,
  title = {Nuclear Configurations in the Spin-Orbit Coupling Model. II. Theoretical Considerations},
  author = {Mayer, Maria Goeppert},
  journal = {Phys. Rev.},
  volume = {78},
  issue = {1},
  pages = {22--23},
  numpages = {0},
  year = {1950},
  month = {Apr},
  publisher = {American Physical Society},
  doi = {10.1103/PhysRev.78.22},
  url = {https://link.aps.org/doi/10.1103/PhysRev.78.22}
}

@article{Vaz1974,
  title = {Systematics of reaction cross sections and interaction barriers for charged particles},
  author = {Vaz, Louis C. and Alexander, John M.},
  journal = {Phys. Rev. C},
  volume = {10},
  issue = {2},
  pages = {464--478},
  numpages = {0},
  year = {1974},
  month = {Aug},
  publisher = {American Physical Society},
  doi = {10.1103/PhysRevC.10.464},
  url = {https://link.aps.org/doi/10.1103/PhysRevC.10.464}
}

@article{Alexander1975,
  title = {Estimation of total reaction cross sections from elastic scattering data},
  author = {Alexander, John M. and Delagrange, H. and Fleury, A.},
  journal = {Phys. Rev. C},
  volume = {12},
  issue = {1},
  pages = {149--155},
  numpages = {0},
  year = {1975},
  month = {Jul},
  publisher = {American Physical Society},
  doi = {10.1103/PhysRevC.12.149},
  url = {https://link.aps.org/doi/10.1103/PhysRevC.12.149}
}

@article{Eyal1976,
  title = {Nuclear size and boundary effects on the fusion barrier of oxygen with carbon},
  author = {Eyal, Y. and Beckerman, M. and Chechik, R. and Fraenkel, Z. and Stocker, H.},
  journal = {Phys. Rev. C},
  volume = {13},
  issue = {4},
  pages = {1527--1535},
  numpages = {0},
  year = {1976},
  month = {Apr},
  publisher = {American Physical Society},
  doi = {10.1103/PhysRevC.13.1527},
  url = {https://link.aps.org/doi/10.1103/PhysRevC.13.1527}
}

@article{Hagino2012,
   title={Subbarrier Fusion Reactions and Many-Particle Quantum Tunneling},
   volume={128},
   ISSN={1347-4081},
   url={http://dx.doi.org/10.1143/PTP.128.1061},
   DOI={10.1143/ptp.128.1061},
   number={6},
   journal={Progress of Theoretical Physics},
   publisher={Oxford University Press (OUP)},
   author={Hagino, K. and Takigawa, N.},
   year={2012},
   month=dec, pages={1061–1106} }

@article{Dasgupta1998,
title = {Barrier distributions as a tool to investigate fusion and fission},
journal = {Nuclear Physics A},
volume = {630},
number = {1},
pages = {78-91},
year = {1998},
note = {Nucleus-Nucleus Collisions},
issn = {0375-9474},
doi = {https://doi.org/10.1016/S0375-9474(97)00745-8},
url = {https://www.sciencedirect.com/science/article/pii/S0375947497007458},
author = {M. Dasgupta and D.J. Hinde and J.R. Leigh and K. Hagino},
}

@article{Rowley1991,
title = {On the “distribution of barriers” interpretation of heavy-ion fusion},
journal = {Physics Letters B},
volume = {254},
number = {1},
pages = {25-29},
year = {1991},
issn = {0370-2693},
doi = {https://doi.org/10.1016/0370-2693(91)90389-8},
url = {https://www.sciencedirect.com/science/article/pii/0370269391903898},
author = {N. Rowley and G.R. Satchler and P.H. Stelson}
}

@article{Wilczy2004,
  title = {Empirical nucleus-nucleus potential deduced from fusion excitation functions},
  author = {Siwek-Wilczy\ifmmode \acute{n}\else \'{n}\fi{}ska, K. and Wilczy\ifmmode \acute{n}\else \'{n}\fi{}ski, J.},
  journal = {Phys. Rev. C},
  volume = {69},
  issue = {2},
  pages = {024611},
  numpages = {10},
  year = {2004},
  month = {Feb},
  publisher = {American Physical Society},
  doi = {10.1103/PhysRevC.69.024611},
  url = {https://link.aps.org/doi/10.1103/PhysRevC.69.024611}
}

@article{Timmers1998,
title = {A case study of collectivity, transfer and fusion enhancement},
journal = {Nuclear Physics A},
volume = {633},
number = {3},
pages = {421-445},
year = {1998},
issn = {0375-9474},
doi = {https://doi.org/10.1016/S0375-9474(98)00121-3},
url = {https://www.sciencedirect.com/science/article/pii/S0375947498001213},
author = {H. Timmers and D. Ackermann and S. Beghini and L. Corradi and J.H. He and G. Montagnoli and F. Scarlassara and A.M. Stefanini and N. Rowley},
}

@article{Scamps2018,
  title = {Examining empirical evidence of the effect of superfluidity on the fusion barrier},
  author = {Scamps, Guillaume},
  journal = {Phys. Rev. C},
  volume = {97},
  issue = {4},
  pages = {044611},
  numpages = {10},
  year = {2018},
  month = {Apr},
  publisher = {American Physical Society},
  doi = {10.1103/PhysRevC.97.044611},
  url = {https://link.aps.org/doi/10.1103/PhysRevC.97.044611}
}

@article{Rehm2016,
  title = {Moments of fusion-barrier distributions},
  author = {Rehm, K. E. and Esbensen, H. and Jiang, C. L. and Back, B. B. and Stefanini, A. M. and Montagnoli, G.},
  journal = {Phys. Rev. C},
  volume = {94},
  issue = {4},
  pages = {044612},
  numpages = {8},
  year = {2016},
  month = {Oct},
  publisher = {American Physical Society},
  doi = {10.1103/PhysRevC.94.044612},
  url = {https://link.aps.org/doi/10.1103/PhysRevC.94.044612}
}

@book{RasmussenWilliams2005,
    author = {Rasmussen, Carl Edward and Williams, Christopher K. I.},
    title = {Gaussian Processes for Machine Learning},
    publisher = {The MIT Press},
    year = {2005},
    month = {11},
    isbn = {9780262256834},
    doi = {10.7551/mitpress/3206.001.0001},
    url = {https://doi.org/10.7551/mitpress/3206.001.0001},
    eprint = {https://direct.mit.edu/book-pdf/2514321/book\_9780262256834.pdf},
}

@article{Shi2019,
author = {Shi, Yuge},
title = {Gaussian Processes, not quite for dummies},
journal = {The Gradient},
year = {2019},
howpublished = {\url{https://thegradient.pub/machine-learning-ancient-japan/ } },
}

@article{Jospin2022,
   title={Hands-On Bayesian Neural Networks—A Tutorial for Deep Learning Users},
   volume={17},
   ISSN={1556-6048},
   url={http://dx.doi.org/10.1109/MCI.2022.3155327},
   DOI={10.1109/mci.2022.3155327},
   number={2},
   journal={IEEE Computational Intelligence Magazine},
   publisher={Institute of Electrical and Electronics Engineers (IEEE)},
   author={Jospin, Laurent Valentin and Laga, Hamid and Boussaid, Farid and Buntine, Wray and Bennamoun, Mohammed},
   year={2022},
   month=may, pages={29–48} }

@article{Blei2017,
  title={Variational inference: A review for statisticians},
  author={Blei, David M and Kucukelbir, Alp and McAuliffe, Jon D},
  journal={Journal of the American statistical Association},
  volume={112},
  number={518},
  pages={859--877},
  year={2017},
  publisher={Taylor \& Francis}
}

@article{Lippmann1991,
    author = {Richard, Michael D. and Lippmann, Richard P.},
    title = {Neural Network Classifiers Estimate Bayesian a posteriori Probabilities},
    journal = {Neural Computation},
    volume = {3},
    number = {4},
    pages = {461-483},
    year = {1991},
    month = {12},
    issn = {0899-7667},
    doi = {10.1162/neco.1991.3.4.461},
    url = {https://doi.org/10.1162/neco.1991.3.4.461},
    eprint = {https://direct.mit.edu/neco/article-pdf/3/4/461/812170/neco.1991.3.4.461.pdf},
}

@article{Hoffman2014,
  author  = {Matthew D. Hoffman and Andrew Gelman},
  title   = {The No-U-Turn Sampler: Adaptively Setting Path Lengths in Hamiltonian Monte Carlo},
  journal = {Journal of Machine Learning Research},
  year    = {2014},
  volume  = {15},
  number  = {47},
  pages   = {1593--1623},
  url     = {http://jmlr.org/papers/v15/hoffman14a.html}
}

@article{Duane1987,
title = {Hybrid Monte Carlo},
journal = {Physics Letters B},
volume = {195},
number = {2},
pages = {216-222},
year = {1987},
issn = {0370-2693},
doi = {https://doi.org/10.1016/0370-2693(87)91197-X},
url = {https://www.sciencedirect.com/science/article/pii/037026938791197X},
author = {Simon Duane and A.D. Kennedy and Brian J. Pendleton and Duncan Roweth},
}

@article{Metropolis1953,
    author = {Metropolis, Nicholas and Rosenbluth, Arianna W. and Rosenbluth, Marshall N. and Teller, Augusta H. and Teller, Edward},
    title = {Equation of State Calculations by Fast Computing Machines},
    journal = {The Journal of Chemical Physics},
    volume = {21},
    number = {6},
    pages = {1087-1092},
    year = {1953},
    month = {06},
}

@Inbook{Neal1996,
author="Neal, Radford M.",
title="Priors for Infinite Networks",
bookTitle="Bayesian Learning for Neural Networks",
year="1996",
publisher="Springer New York",
address="New York, NY",
pages="29--53",
isbn="978-1-4612-0745-0",
doi="10.1007/978-1-4612-0745-0_2",
url="https://doi.org/10.1007/978-1-4612-0745-0_2"
}

@misc{Pearce2019,
      title={Expressive Priors in Bayesian Neural Networks: Kernel Combinations and Periodic Functions}, 
      author={Tim Pearce and Russell Tsuchida and Mohamed Zaki and Alexandra Brintrup and Andy Neely},
      year={2019},
      eprint={1905.06076},
      archivePrefix={arXiv},
      primaryClass={stat.ML},
      url={https://arxiv.org/abs/1905.06076}, 
}

@inproceedings{
Meronen2021,
title={Periodic Activation Functions Induce Stationarity},
author={Lassi Meronen and Martin Trapp and Arno Solin},
booktitle={Advances in Neural Information Processing Systems},
editor={A. Beygelzimer and Y. Dauphin and P. Liang and J. Wortman Vaughan},
year={2021},
url={https://openreview.net/forum?id=gRwh5HkdaTm}
}

@phdthesis{Duvenaud2014, 
title={Automatic model construction with Gaussian processes}, url={https://www.repository.cam.ac.uk/handle/1810/247281}, DOI={10.17863/CAM.14087}, school={Apollo - University of Cambridge Repository}, author={Duvenaud, David}, year={2014} }

@article{Morton1999,
  title = {Coupled-channels analysis of the ${}^{16}\mathrm{O}{+}^{208}\mathrm{Pb}$ fusion barrier distribution},
  author = {Morton, C. R. and Berriman, A. C. and Dasgupta, M. and Hinde, D. J. and Newton, J. O. and Hagino, K. and Thompson, I. J.},
  journal = {Phys. Rev. C},
  volume = {60},
  issue = {4},
  pages = {044608},
  numpages = {11},
  year = {1999},
  month = {Aug},
  publisher = {American Physical Society},
  doi = {10.1103/PhysRevC.60.044608},
  url = {https://link.aps.org/doi/10.1103/PhysRevC.60.044608}
}

@article{Montagnoli2012,
  title = {Fusion of ${}^{\text{40}}\text{Ca}+{}^{\text{40}}\text{Ca}$ and other $\text{Ca}+\text{Ca}$ systems near and below the barrier},
  author = {Montagnoli, G. and Stefanini, A. M. and Jiang, C. L. and Esbensen, H. and Corradi, L. and Courtin, S. and Fioretto, E. and Goasduff, A. and Haas, F. and Kifle, A. F. and Michelagnoli, C. and Montanari, D. and Mijatovi\ifmmode \acute{c}\else \'{c}\fi{}, T. and Rehm, K. E. and Silvestri, R. and Singh, Pushpendra P. and Scarlassara, F. and Szilner, S. and Tang, X. D. and Ur, C. A.},
  journal = {Phys. Rev. C},
  volume = {85},
  issue = {2},
  pages = {024607},
  numpages = {8},
  year = {2012},
  month = {Feb},
  publisher = {American Physical Society},
  doi = {10.1103/PhysRevC.85.024607},
  url = {https://link.aps.org/doi/10.1103/PhysRevC.85.024607}
}

@article{Jiang2010,
  title = {Fusion hindrance for Ca$+$Ca systems: Influence of neutron excess},
  author = {Jiang, C. L. and Stefanini, A. M. and Esbensen, H. and Rehm, K. E. and Corradi, L. and Fioretto, E. and Mason, P. and Montagnoli, G. and Scarlassara, F. and Silvestri, R. and Singh, P. P. and Szilner, S. and Tang, X. D. and Ur, C. A.},
  journal = {Phys. Rev. C},
  volume = {82},
  issue = {4},
  pages = {041601},
  numpages = {4},
  year = {2010},
  month = {Oct},
  publisher = {American Physical Society},
  doi = {10.1103/PhysRevC.82.041601},
  url = {https://link.aps.org/doi/10.1103/PhysRevC.82.041601}
}

@article{Stefanini2009,
title = {How does fusion hindrance show up in medium-light systems? The case of 48Ca+48Ca},
journal = {Physics Letters B},
volume = {679},
number = {2},
pages = {95-99},
year = {2009},
issn = {0370-2693},
doi = {https://doi.org/10.1016/j.physletb.2009.07.017},
url = {https://www.sciencedirect.com/science/article/pii/S0370269309008351},
author = {A.M. Stefanini and G. Montagnoli and R. Silvestri and L. Corradi and S. Courtin and E. Fioretto and B. Guiot and F. Haas and D. Lebhertz and P. Mason and F. Scarlassara and S. Szilner},
}

@article{Aljuwair1984,
  title = {Isotopic effects in the fusion of $^{40}\mathrm{Ca}$ with $^{40,44,48}\mathrm{Ca}$},
  author = {Aljuwair, H. A. and Ledoux, R. J. and Beckerman, M. and Gazes, S. B. and Wiggins, J. and Cosman, E. R. and Betts, R. R. and Saini, S. and Hansen, Ole},
  journal = {Phys. Rev. C},
  volume = {30},
  issue = {4},
  pages = {1223--1227},
  numpages = {0},
  year = {1984},
  month = {Oct},
  publisher = {American Physical Society},
  doi = {10.1103/PhysRevC.30.1223},
  url = {https://link.aps.org/doi/10.1103/PhysRevC.30.1223}
}

@article{Trotta2001,
  title = {Sub-barrier fusion of the magic nuclei ${}^{40,48}\mathrm{Ca}{+}^{48}\mathrm{Ca}$},
  author = {Trotta, M. and Stefanini, A. M. and Corradi, L. and Gadea, A. and Scarlassara, F. and Beghini, S. and Montagnoli, G.},
  journal = {Phys. Rev. C},
  volume = {65},
  issue = {1},
  pages = {011601},
  numpages = {5},
  year = {2001},
  month = {Dec},
  publisher = {American Physical Society},
  doi = {10.1103/PhysRevC.65.011601},
  url = {https://link.aps.org/doi/10.1103/PhysRevC.65.011601}
}

@article{Dasgputa2007,
  title = {Beyond the Coherent Coupled Channels Description of Nuclear Fusion},
  author = {Dasgupta, M. and Hinde, D. J. and Diaz-Torres, A. and Bouriquet, B. and Low, Catherine I. and Milburn, G. J. and Newton, J. O.},
  journal = {Phys. Rev. Lett.},
  volume = {99},
  issue = {19},
  pages = {192701},
  numpages = {4},
  year = {2007},
  month = {Nov},
  publisher = {American Physical Society},
  doi = {10.1103/PhysRevLett.99.192701},
  url = {https://link.aps.org/doi/10.1103/PhysRevLett.99.192701}
}

@article{Yukawa1935,
    author = "Yukawa, Hideki",
    title = "{On the Interaction of Elementary Particles I}",
    reportNumber = "PRINT-92-0144",
    doi = "10.1143/PTPS.1.1",
    journal = "Proc. Phys. Math. Soc. Jap.",
    volume = "17",
    pages = "48--57",
    year = "1935"
}

@article{Lattes1947,
  author    = {Lattes, C. M. G. and Muirhead, H. and Occhialini, G. P. S. and Powell, C. F.},
  title     = {Processes Involving Charged Mesons},
  journal   = {Nature},
  volume    = {159},
  pages     = {694--697},
  year      = {1947}
}

@article{GellMann1964,
  author    = {Gell-Mann, Murray},
  title     = {A Schematic Model of Baryons and Mesons},
  journal   = {Physics Letters},
  volume    = {8},
  pages     = {214--215},
  year      = {1964}
}

@article{Zweig1964,
  author    = {Zweig, G.},
  title     = {An SU(3) Model for Strong Interaction Symmetry and Its Breaking, Version 2},
  journal   = {CERN preprint},
  number    = {TH-401},
  year      = {1964}
}

@article{Friedman1969,
  author    = {Friedman, J. I. and Kendall, H. W. and Taylor, R. E.},
  title     = {Deep Inelastic Scattering: Experimental Evidence for Quark Constituents in the Proton},
  journal   = {Physical Review Letters},
  volume    = {23},
  pages     = {930--934},
  year      = {1969}
}

@article{Gerjuoy1942,
  title = {On Tensor Forces and the Theory of Light Nuclei},
  author = {Gerjuoy, Edward and Schwinger, Julian},
  journal = {Phys. Rev.},
  volume = {61},
  issue = {3-4},
  pages = {138--146},
  numpages = {0},
  year = {1942},
  month = {Feb},
  publisher = {American Physical Society},
  doi = {10.1103/PhysRev.61.138},
  url = {https://link.aps.org/doi/10.1103/PhysRev.61.138}
}

@article{Aprile2019,
  author = {Aprile, E. and others},
  title = {Observation of two-neutrino double electron capture in $^{124}$Xe with XENON1T},
  journal = {Nature},
  volume = {568},
  pages = {532--535},
  year = {2019},
  doi = {10.1038/s41586-019-1124-4}
}

@article{Huxford2024,
  title = {Accuracy of neutron star radius measurement with the next generation of terrestrial gravitational-wave observatories},
  author = {Huxford, Rachael and Kashyap, Rahul and Borhanian, Ssohrab and Dhani, Arnab and Gupta, Ish and Sathyaprakash, B. S.},
  journal = {Phys. Rev. D},
  volume = {109},
  issue = {10},
  pages = {103035},
  numpages = {26},
  year = {2024},
  month = {May},
  publisher = {American Physical Society},
  doi = {10.1103/PhysRevD.109.103035},
  url = {https://link.aps.org/doi/10.1103/PhysRevD.109.103035}
}

@article{Kelkar2016,
   title={Extraction of the Proton Charge Radius from Experiments},
   volume={20},
   ISSN={2339-1995},
   url={http://dx.doi.org/10.7454/mss.v20i3.6242},
   DOI={10.7454/mss.v20i3.6242},
   number={3},
   journal={Makara Journal of Science},
   publisher={Universitas Indonesia, Directorate of Research and Public Service},
   author={Kelkar, Neelima Govind and Mart, Terry and Nowakowski, Marek},
   year={2016},
   month=sep }

@article{Sun2025,
  author = {Sun, Z. H. and Ekstr\"om, A. and Forss\'en, C. and Hagen, G. and Jansen, G. R. and Papenbrock, T.},
  title = {Multiscale Physics of Atomic Nuclei from First Principles},
  journal = {Phys. Rev. X},
  volume = {15},
  number = {1},
  pages = {011028},
  year = {2025},
  doi = {10.1103/PhysRevX.15.011028}
}

@article{Kondev2021,
doi = {10.1088/1674-1137/abddae},
url = {https://dx.doi.org/10.1088/1674-1137/abddae},
year = {2021},
month = {mar},
publisher = {Chinese Physical Society and the Institute of High Energy Physics of the Chinese Academy of Sciences and the Institute of Modern Physics of the Chinese Academy of Sciences and IOP Publishing Ltd},
volume = {45},
number = {3},
pages = {030001},
author = {Kondev, F.G. and Wang, M. and Huang, W.J. and Naimi, S. and Audi, G.},
title = {The NUBASE2020 evaluation of nuclear physics properties *},
journal = {Chinese Physics C},}

@article{Fernbach1955,
  author = {Fernbach, S. and Heckrotte, W. and Lepore, J.V.},
  title = {Theory of Polarization of Nucleons Scattered Elastically by Nuclei},
  journal = {Phys. Rev.},
  volume = {97},
  pages = {1059--1071},
  year = {1955},
  doi = {10.1103/PhysRev.97.1059}
}

@article{Bethe1949,
  title = {Theory of the Effective Range in Nuclear Scattering},
  author = {Bethe, H. A.},
  journal = {Phys. Rev.},
  volume = {76},
  issue = {1},
  pages = {38--50},
  numpages = {0},
  year = {1949},
  month = {Jul},
  publisher = {American Physical Society},
  doi = {10.1103/PhysRev.76.38},
  url = {https://link.aps.org/doi/10.1103/PhysRev.76.38}
}

@article{Kellogg1939,
  title = {An Electrical Quadrupole Moment of the Deuteron},
  author = {Kellogg, J. M. B. and Rabi, I. I. and Ramsey, N. F. and Zacharias, J. R.},
  journal = {Phys. Rev.},
  volume = {55},
  issue = {3},
  pages = {318--319},
  numpages = {0},
  year = {1939},
  month = {Feb},
  publisher = {American Physical Society},
  doi = {10.1103/PhysRev.55.318},
  url = {https://link.aps.org/doi/10.1103/PhysRev.55.318}
}

@article{Giuliani2024,
  title = {Model orthogonalization and Bayesian forecast mixing via principal component analysis},
  author = {Giuliani, P. and Godbey, K. and Kejzlar, V. and Nazarewicz, W.},
  journal = {Phys. Rev. Res.},
  volume = {6},
  issue = {3},
  pages = {033266},
  numpages = {9},
  year = {2024},
  month = {Sep},
  publisher = {American Physical Society},
  doi = {10.1103/PhysRevResearch.6.033266},
  url = {https://link.aps.org/doi/10.1103/PhysRevResearch.6.033266}
}

@misc{AutoBNN2024,
  title        = {{AutoBNN: Probabilistic Time Series Forecasting with Compositional Bayesian Neural Networks}},
  author       = {Carroll, Colin and Colthurst, Thomas and K{\"o}ster, Urs and Vasudevan, Srinivas},
  year         = {2024},
  howpublished = {GitHub repository, \url{https://github.com/tensorflow/probability/tree/main/spinoffs/autobnn}},
  note         = {Open-source software (Google Research Blog announcement)}
}

@article{HohenbergKohn1964,
  author    = {Hohenberg, Pierre and Kohn, Walter},
  title     = {Inhomogeneous Electron Gas},
  journal   = {Physical Review},
  volume    = {136},
  pages     = {B864--B871},
  year      = {1964}
}

@article{KohnSham1965,
  author    = {Kohn, Walter and Sham, Lu Jeu},
  title     = {Self-Consistent Equations Including Exchange and Correlation Effects},
  journal   = {Physical Review},
  volume    = {140},
  pages     = {A1133--A1138},
  year      = {1965}
}

@article{VautherinBrink1972,
  author    = {Vautherin, Denis and Brink, David M.},
  title     = {Hartree--Fock Calculations with Skyrme's Interaction. I. Spherical Nuclei},
  journal   = {Physical Review C},
  volume    = {5},
  pages     = {626--647},
  year      = {1972}
}

@article{Gogny1980,
  title = {Hartree-Fock-Bogolyubov calculations with the $D1$ effective interaction on spherical nuclei},
  author = {Decharg\'e, J. and Gogny, D.},
  journal = {Phys. Rev. C},
  volume = {21},
  issue = {4},
  pages = {1568--1593},
  numpages = {0},
  year = {1980},
  month = {Apr},
  publisher = {American Physical Society},
  doi = {10.1103/PhysRevC.21.1568},
  url = {https://link.aps.org/doi/10.1103/PhysRevC.21.1568}
}

@article{Negele1972,
  title = {Density-Matrix Expansion for an Effective Nuclear Hamiltonian},
  author = {Negele, J. W. and Vautherin, D.},
  journal = {Phys. Rev. C},
  volume = {5},
  issue = {5},
  pages = {1472--1493},
  numpages = {0},
  year = {1972},
  month = {May},
  publisher = {American Physical Society},
  doi = {10.1103/PhysRevC.5.1472},
  url = {https://link.aps.org/doi/10.1103/PhysRevC.5.1472}
}

@misc{Philip2026,
  author       = {Philip, Aaron},
  title        = {barrier-extraction},
  year         = {2026},
  howpublished = {\url{https://github.com/aaron-philip/barrier-extraction}},
  note         = {GitHub repository, accessed June 29, 2026}
}

@book{casella2024,
  title={Statistical inference},
  author={Casella, George and Berger, Roger},
  year={2024},
  publisher={Chapman and Hall/CRC}
}

@article{gneiting2007,
  title={Strictly proper scoring rules, prediction, and estimation},
  author={Gneiting, Tilmann and Raftery, Adrian E},
  journal={J. Am. Stat. Assoc.},
  volume={102},
  number={477},
  pages={359--378},
  year={2007},
  publisher={Taylor \& Francis},
  doi = {10.1198/016214506000001437},
}

\end{document}